\documentclass[sigconf]{acmart} 
\renewcommand\footnotetextcopyrightpermission[1]{}

\AtBeginDocument{%
  \providecommand\BibTeX{{%
    \normalfont B\kern-0.5em{\scshape i\kern-0.25em b}\kern-0.8em\TeX}}}

\setcopyright{acmcopyright}
\copyrightyear{2018}
\acmYear{2018}
\acmDOI{XXXXXXX.XXXXXXX}

\acmConference[Conference acronym 'XX]{Make sure to enter the correct
  conference title from your rights confirmation emai}{June 03--05,
  2018}{Woodstock, NY}
\acmPrice{15.00}
\acmISBN{978-1-4503-XXXX-X/18/06}

\usepackage{subcaption}
\usepackage{longtable}
\begin{document}

\title{Software Update Practices on Smart Home IoT Devices}

\author{Vijay Prakash}
\affiliation{%
  \institution{New York University}
  \city{New York}
  \country{USA}}
\email{vijay.prakash@nyu.edu}

\author{Sicheng Xie}
\affiliation{%
  \institution{New York University}
  \city{New York}
  \country{USA}}
\email{sx810@nyu.edu}

\author{Danny Yuxing Huang}
\affiliation{%
  \institution{New York University}
  \city{New York}
  \country{USA}}
\email{dhuang@nyu.edu}

\begin{abstract}
Smart home IoT devices are known to be breeding grounds for security and privacy vulnerabilities. Although some IoT vendors deploy updates, the update process is mostly opaque to researchers. It is unclear what software components are on devices, whether and when these components are updated, and how vulnerabilities change alongside the updates. This opaqueness makes it difficult to understand the security of software supply chains of IoT devices.

To understand the software update practices on IoT devices, we leverage IoT Inspector's dataset of network traffic from real-world IoT devices. We analyze the User Agent strings from plain-text HTTP connections. We focus on four software components included in User Agents: cURL, Wget, OkHttp, and python-requests. By keeping track of what kinds of devices have which of these components at what versions, we find that many IoT devices potentially used outdated and vulnerable versions of these components---based on the User Agents---even though less vulnerable, more updated versions were available; and that the rollout of updates tends to be slow for some IoT devices. 
\end{abstract}

\ccsdesc[500]{Security and privacy~Mobile and wireless security}
\ccsdesc[100]{General and reference~Empirical studies}

\keywords{IoT; supply chain; updates}

\maketitle
\pagestyle{plain}
\settopmatter{printfolios=true} %
\renewcommand{\paragraph}[1]{\vspace{0.1cm} \noindent \textbf{#1}}
\newcommand{\XXX}[0]{{\color{red} XXX}}
\newcommand{\danny}[1]{{\color{blue} #1}}
\newcommand{\vijay}[1]{{\color{purple} {#1}{-- Vijay}}}

\section{Introduction}

Smart home technologies, also known as smart devices or Internet-of-Things (IoT) devices, are gaining popularity, such
as smart TVs, speakers, cameras, and medical devices, yet they are breeding grounds for security and privacy
threats~\cite{usenix_mirai,kumar2019all}, using vulnerable software components~\cite{TLSFingerprints} and putting the
users and other hosts on the network at risk. 

To mitigate these risks, like a software, many IoT devices offer updates that can be installed automatically or manually by users. 
However, this update process remains largely opaque; it is unclear what software components are on devices, whether/how
any software updates have occurred on these software components, and how vulnerabilities change alongside the updates
across a large number, variety, and the long tail of IoT devices in smart homes~\cite{IoTCompanies}. This opaqueness
makes it hard to keep track of the software bill-of-materials (SBOMs) across devices over time and understand the
software supply chain of smart home IoT devices in general.

One of the reasons for this knowledge gap is the lack of scale in many lab-based analyses of IoT devices. To understand
the software components on IoT devices and the update behaviors, researchers often analyze the firmware binaries.
However, commercial smart home IoT devices tend to use  proprietary and/or protected software components (such as the
firmware and libraries) that are difficult to extract and reverse-engineer~\cite{mohajeri2019watching}. Furthermore,
researchers often need the physical IoT devices for their analyses. The number of devices that can be studied is often
constrained by time and budget; one of the largest sample of IoT devices studied in the lab is about 120
devices~\cite{ren2019information,paracha2021iotls} and does not cover many devices in the long tail~\cite{IoTCompanies}.
As such, it is difficult for researchers to identify whether and how IoT devices update to patch vulnerabilities over time. 

\begin{figure}[!h]
    \centering \includegraphics[width=\linewidth,trim={1.0cm 1.0cm 1.0cm 9.5cm},clip,scale=0.60]{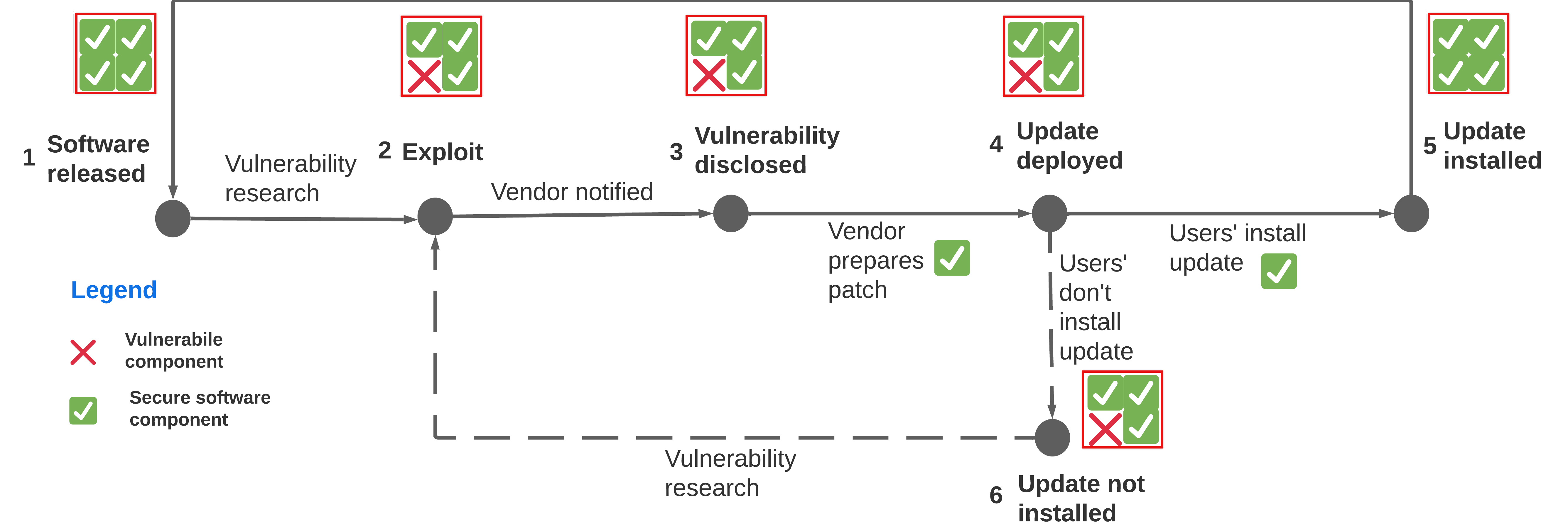}
    \caption{Software through vulnerability cycle. In this study, we are focusing on the updates, step 4, 5 and 6.}
    \label{fig:vuln-cycle} 
\end{figure}

Figure \ref{fig:vuln-cycle} shows software through vulnerability life cycle: 1) A software is released; 2) Security researchers find vulnerabilities; 3) Researchers ethically notify vendors about the vulnerabilities privately and give a grace period of 90 days (current standard) to fix those issues and release updates; 4) Vendor publicly discloses the vulnerability along with updates containing fixes; 5) In the ideal case, users download the updates and update the software making it secure against the exploits; 6) In the worst case, users don’t install the update and software stays vulnerable. Finally the cycle continues starting from vulnerability research again after both the 5th and 6th stage of the cycle.

\paragraph{Objectives} Our goal is to understand software update practices on IoT devices in real-world smart homes:
what software components and versions are present on devices at a particular time, as well as when the versions and the
associated vulnerabilities have changed over time (i.e., likely updates). This
knowledge will shed light on the update practices across a large number and variety of IoT devices and vendors, thus
paving the way for deeper empirical understanding of IoT software supply chain security. 

\paragraph{Method} To achieve this goal, we conduct
the first known longitudinal analysis of IoT software components in real-world smart homes and provide an initial
understanding of update practices in the IoT ecosystem. In particular, we leverage IoT Inspector's dataset of network
traffic crowdsourced from smart homes around the world~\cite{danny-01}. The dataset includes the likely identities of
devices, along with any User-Agent strings extracted from plain-text HTTP headers sent by these devices. The User Agents
indicate \textit{potential} use of certain software components and their versions. We would like to note that devices whose HTTP traffic we have captured do not use plain HTTP for the majority of communication, but they do send them out for some reason with empty payload, possibly to test a connection or discovery purposes (based on our lab studies). We have not decrypted the HTTPS connections while collecting data for this study.

In this paper, we analyze the versions of four different software components that appear in the User Agent strings:
cURL (which is both a library, \texttt{libcurl}, and a binary), Wget (a binary), OkHttp library, and Requests (Python's)
library. We observe these components in the User Agent strings on 23,837 devices (including 359 what looks like IoT
devices) across 4,562 real-world smart homes between Apr 2019 and Oct 2021.

\paragraph{Findings.} We show that update practices in the IoT ecosystem are different from (sometimes worse than)
findings from general purpose computing devices.  In particular, none of the IoT devices in our dataset included the
latest versions of the four software components in the User Agent, even when the latest versions were available at the
time IoT Inspector captured the data. In some cases, the lower versions included critical CVEs that could have been
fixed if the IoT devices had been using the latest versions at the time.  Furthermore, we find that vendor deploy
updates in rolling fashion, and oftentimes rolled out updates are not the latest. In other words, when vendors deploy
updates, they do not always update the software component to the latest versions, which means oftentimes devices are
left in vulnerable state even after end user install the update.

These findings paint a grim picture of the slow (and sometimes the lack of) software update practices on certain IoT
devices. Such update practices could be a result of users not promptly installing updates, IoT vendors not deploying
updates in time, or both---it is an open question what is the cause for our observation. It is our hope that this study
will offer the first evidence on the update practices on IoT devices, highlight the associated software supply chain
risks, and provide the impetus for more secure practices on users, IoT vendors, and regulators.

\section{Related Work}

\paragraph{Software updates.}
In the last couple of decades, there has been a push from academia and industry to improve the software update hygiene for all kinds of software because old software has been correlated with compromise \cite{bilge-01,khan-01}, so it is in the best interest of everyone's security and from economical standpoint to use the latest version of software \cite{louis-01,sarabi-01, arbaugh-01, sandy-01, philipp-01}. Propelled by this push, major software vendors have adopted the practice of deploying frequent updates, which improved the vendor update deployment practice. On the other hand, end users are advised to follow “Best Practices”, which includes updating software as soon as updates are available if users can \cite{bilge-01,khan-01,sarabi-01}. For software like operating systems (OS) and applications running on them, previous studies have shown that user update installation practice can be modeled with either geometric or exponential distribution, which means majority of the users update the software after new release, with some users taking longer duration to install the updates \cite{sarabi-01,eric-01,vuln-law-1,vuln-law-2}. 

Updating software is considered to be a part of "Best Practices" to increase security by shrinking the vulnerable state of hosts, \cite{robert-01,louis-01} recommend updates. Khan et al. \cite{khan-01} found a positive correlation between infection indicators and a lack of regular updating practice, \cite{bilge-01} showed correlation in devices not updating and being compromised at some point of time. Sarabi et al.\cite{sarabi-01} found that frequent discovery of vulnerabilities in a software limits the benefit of providing faster updates. DeKoven et al. \cite{louis-01} measure the correlation between ``Best Practices'' and impact on security risk. 

\paragraph{Four factors for update.}
One way to define security of a host is how vulnerable it is; the less vulnerability a host has, the more secure it likely is. From the software update perspective, there are four factors that decide the vulnerable state of a host: 1) how quickly users install the updates; 2) how quickly vendors deploy the patch; 3) how vendors deploy the patch; and 4) how frequently vulnerabilities are found \cite{sarabi-01}. These combined together tell about update practice. 
So far studies have looked at all the four factors for software that are used for computing in general (i.e. mobiles, computers, tablets, etc.) \cite{alhazmi-02,sandy-01,arbaugh-01,cavusoglu-01,alhazmi-01, arora-01, neuhaus-01, eric-02, shahzad-01, zakir-01, antonio-01, vuln-law-1, eric-01, scott-01, sarabi-01}. 

There have been studies about all the factors affecting the vulnerability state of a host. These studies have been conducted for both clients side (general purpose computing devices) and server side hosts. For general purpose computing devices, there is \cite{sarabi-01, louis-01} on user update behaviors using various software; \cite{antonio-01, arbaugh-01, sandy-01, alhazmi-01, alhazmi-02, arora-01, neuhaus-01, eric-02, shahzad-01} on
vendor update deployment and vulnerability disclosure; \cite{christos-01, stefan-01, thomas-01} on mechanism of deploying patches where they suggest silent updates are the most effective.
On servers side hosts update behaviors there are \cite{vuln-law-1, vuln-law-2, eric-01, zakir-01, scott-01}, where they measure patch installation delays after major security incidents and routine patch releases after vulnerability disclosures.

\paragraph{Update practice for IoT.}
Specific to  IoT devices, Yousefnezhad et al.\cite{narges-01} has done a comprehensive survey of IoT product lifecycle, including vulnerability management and software updates. There have been numerous study about IoT update infrastructure, i.e., how to deliver updates to IoT devices logistically \cite{alexander-01, kim-01, chi-01, langiu-01}, and securely \cite{meriem-01}. It has also been found that user awareness about updating devices has improved — IoT device users prefer to buy products that guarantees updates \cite{philipp-01}. 

So far we are not aware of any study about the current update practice in IoT ecosystem.
The case for IoT devices is different from major software vendors because of the sheer amount of devices and vendors available in the market~\cite{IoTCompanies}. Another possible reason for the differentiation could be that vendor nudges to update software could not reach IoT device users, unlike computing devices if users are actively using devices they see the update nudge on UI. Because some IoT devices can only be interacted with companion apps, and users might forget to use the apps as these devices might be always on without user interaction. Also, unlike few major software vendors for personal computing, IoT vendors are fragmented and too many~\cite{IoTCompanies}. All this makes it difficult to apply the results of previous studies to the IoT ecosystem.

\vspace{-10.0bp}
\section{Method}

\subsection{Inferring software components from User Agents}
To understand software update practices on real-world IoT devices, we use IoT Inspector's \cite{danny-01} dataset of network traffic, crowdsourced from global smart home users. The dataset includes names and manufacturers of devices, along with various network header data, including the User Agent strings extracted from plain-text HTTP connections. Note that IoT Inspector does not capture the contents of the HTTP traffic; nor does it collect personally identifiable information~\cite{danny-01}. We look at a subset of this data, which consists of 4,562 different smart homes containing 23,837 unique devices (including IoT and general purpose computing)  from Apr 9, 2019 to Oct 12, 2021.

The captured User Agent strings offer indications for software components that devices have \textit{likely} used. We focus on four software components used as User Agents: cURL \cite{cURL}, Wget \cite{Wget}, Python Requests \cite{requests}, and OkHttp \cite{okhttp}.  We picked these because they are the most popular (by the number of devices) User Agents that are not browsers. 
This information is by no means the ground truth for what software components devices \textit{actually} used (since HTTP clients can, in theory, put anything into the User Agent field). Still, we assume that devices' HTTP clients are honest about being non-browsers. There is evidence for non-truthful User Agents for browsers~\cite{UA-Lying}, but we are unaware of any evidence of non-truthful User Agents for non-browsers like cURL, Wget, Requests, and OkHttp. As such, throughout this paper, we use the User Agent as a proxy for the actual software components. 

IoT Inspector captures the HTTP User-Agent header seen in the network traffic from uniquely identified devices with a timestamp of when it was seen. Devices whose traffic is captured could include general purpose computing devices (computers, phones, etc.) and IoT devices. We distinguish the device categories with the Fingerbank \cite{fingerbank} APIs (which identify devices based on the MAC address and destination hosts contacted over a proprietary ML model).

To find out how old and vulnerable versions of these four software are used, we manually compile a list of versions released with dates from their websites or code bases. Each version is assigned a release number from 0 (first release) to N (last release). We also compile a table of vulnerabilities each version of these four software have. Vulnerabilities in each version are either available on their website or we compile it by searching names of each software on the National Vulnerability Database (NVD) \cite{nvd} and use the CVE number and Common Platform Enumeration (CPE) to map them to different versions.

\vspace{-0.5em}
\subsection{Metrics for understanding update practices}
To get the understanding of update practices, from the collected data where HTTP user agent is one of the four software, and compilation of four software version release dates and vulnerability each version have, we curate statistical metrics for all the unique device and user agents combination. This includes following: 
\begin{itemize}
    \item What software is used as user agent and its used version
    \item First and last seen timestamp in the IoT Inspector dataset
    \item At the last seen time, the latest available version of that software 
    \item At the last seen time, the number of versions and days (or months) behind this used user agent version is from this latest available version at this time
    \item At the last seen time, the number of vulnerabilities that could have been avoided by using the latest version. We explain how this is calculated later in the next subsection \ref{metric-explan}. 
\end{itemize}
\noindent
For our analysis to gain perspective of update practices we rely on three metrics retroactively: 

\begin{enumerate}
    \item number of avoidable vulnerabilities, having vulnerabilities directly contribute to the vulnerable state of a host adversely
    \item number of versions behind
    \item and number of days behind (directly correlated with number of versions behind), they both show how quickly hosts are updated which involves both user promptness in installing updates and vendors deploying updates.
\end{enumerate}
\noindent
From our data set we can't distinguish if users are not installing updates promptly or vendors are not deploying updates, but we can tell about the updates practices comprising both with these three metrics. 

\subsection{How these metrics are calculated} \label{metric-explan}

Versions behind for a user agent when it's seen in our data is calculated by subtracting the release number of the used version from release number of most recent available version at this seen time. To calculate the number of avoidable vulnerabilities, we take the difference between the set of all the vulnerabilities in the used version and the set of vulnerabilities in the most recent version, which tells us vulnerabilities that were in the used version but not in the latest version, i.e., they were fixed in the latest version. Let's say, X be the set of vulnerabilities in the used version and Y be the set of vulnerabilities in the latest available version. Number of avoidable vulnerabilities would be (X - Y), whereas this - (subtraction) represents the difference of set operation. 

\begin{figure}[!b]
    \centering
    \includegraphics[trim={0 0 0 2.6cm},clip,scale=0.35]{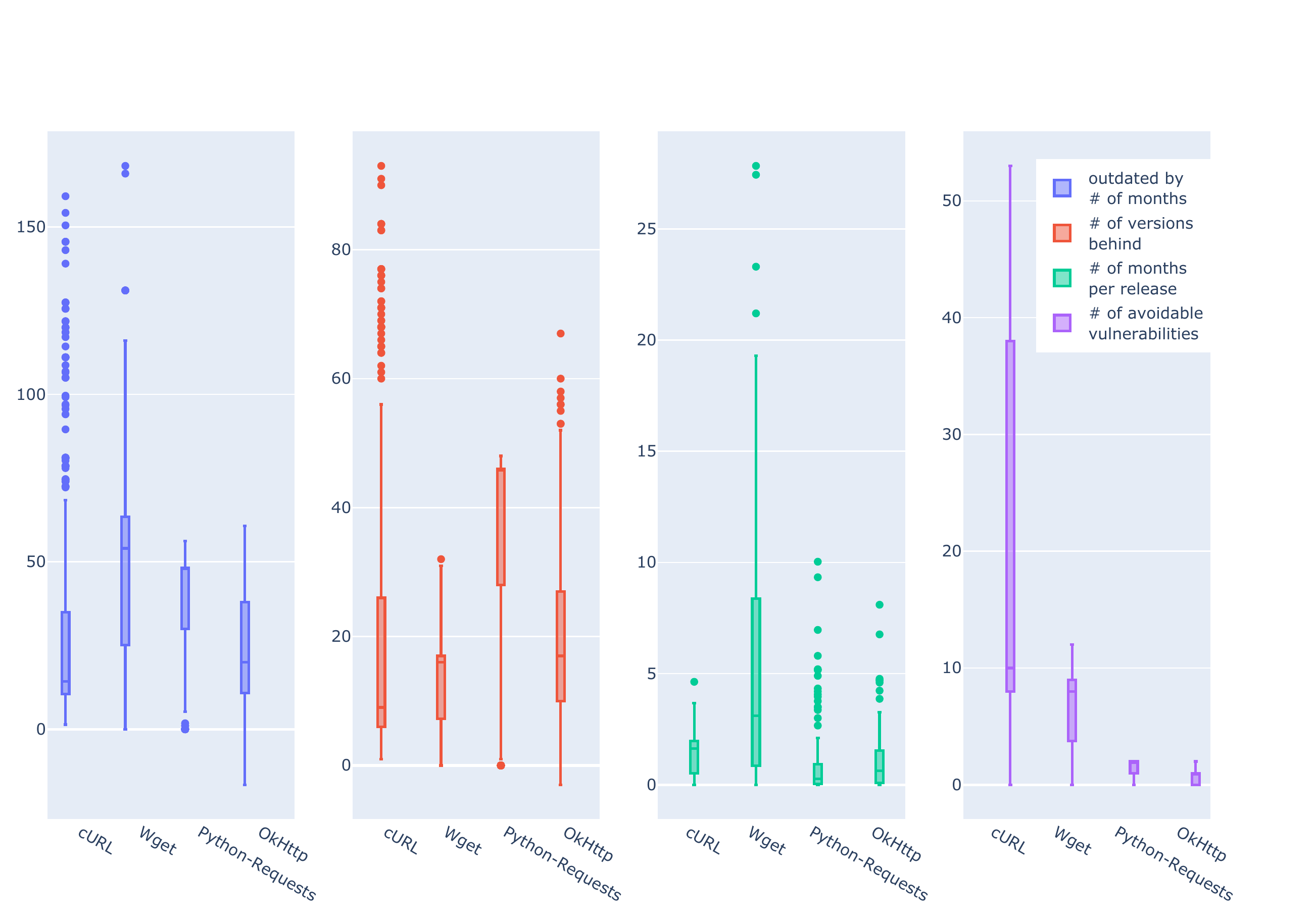}
    \setlength{\abovecaptionskip}{-10pt}
    \setlength{\belowcaptionskip}{-10pt}
    \caption{Distribution of different metrics for all software components in our data.}
    \label{fig:all-ua-distr}
\end{figure}

\begin{figure*}[!ht]
    \centering
    \includegraphics[trim={0 0 0 2.6cm},clip,scale=0.4]{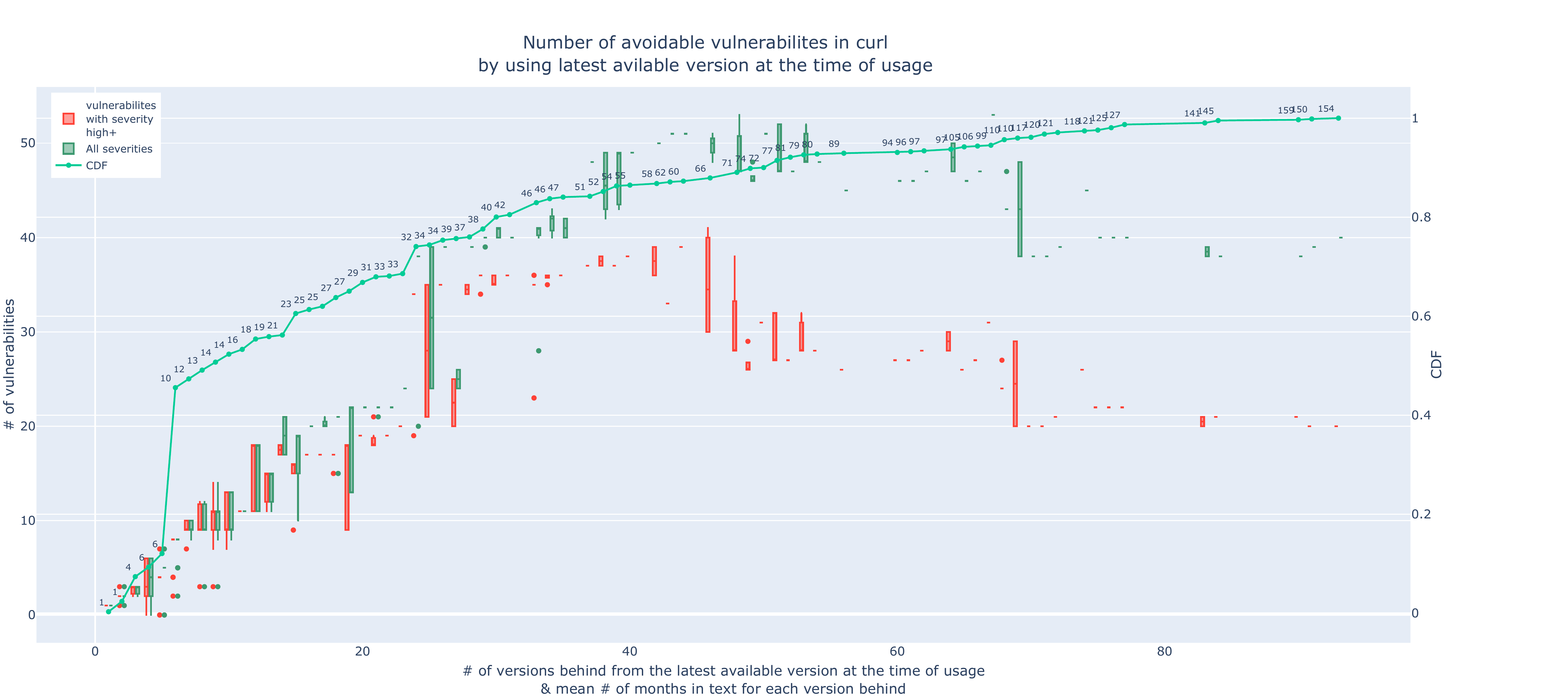}
    \setlength{\abovecaptionskip}{-10pt}
    \setlength{\belowcaptionskip}{-10pt}
    \caption{Number of avoidable vulnerabilities in cURL against number of versions behind, and CDF distribution of devices for number of versions behind.}
    \label{fig:curl-vuln-version}
\end{figure*}

For example, let's suppose cURL version 7.20.0, which is release \# 112, released on 2010-02-09, and has 10 vulnerabilities as of today. Let's also suppose we see a device with the user agent cURL 7.20.0 on 2020-02-09 in the dataset. Latest available version of cURL on 2020-02-09 is version 7.68.0, which is release \# 187, released on 2020-01-08, and has 8 vulnerabilities as of today. The number of versions behind for this seen version is 75 and the number of days behind is 3620 (9 years and 11 months). For the sake of this example, let's assume 8 of the vulnerabilities in 7.20.0 are not present in 7.68.0 and 2 of them were still present on 2020-02-09, which also means 7.68.0 has 6 new vulnerabilities introduced in it but at some point after its release maybe even after the date 2020-02-09. So, the number of vulnerabilities that could have been avoided by using 7.68.0 on 2020-02-09 is 8 because they are fixed in this version. 

\paragraph{Our data set.}Overview of our collected data through these metrics is in the Figure \ref{fig:all-ua-distr}. First, second, and fourth subplots show the comparison of three metrics we described earlier. Third subplot gives us the idea of how frequently updates are released for software components. Looking at these plots all together gives us an complete idea of the dynamic of update practice for these software components, which individual plots don’t show. For e.g., if we just look at the outdated by number of months, we could assume that the order of versions of software components used from more outdated to less outdated would be: cURL, Wget, OkHttp, and Requests. If we associate outdatedness with a vulnerable state, the order of software components would be the same. If we only look at the number of versions behind, the order of vulnerable states would be cURL, OkHttp, Requests, and Wget. Although, this is not true because Wget releases new versions less frequently than all of the other software components. The more frequently a software component releases new versions; it is more likely that versions behind of the used software components are going to be high in the wild. Hence we can’t use only outdatedness by month or versions. Number of avoidable vulnerabilities shows the true measure of how vulnerable a software component is. Using that for our data, performance of four software components from least secure to most secure is: cURL, Wget, Python requests, and OkHTTP. Another thing to note is that in the first and second sublot we could also see that some of the versions reported by IoT devices for OkHttp are wrong as the number of days and versions behind is negative because they were released in the future after they were used in the wild. We looked further into this case and there is only one instance of false reporting in our data.

\vspace{-0.5em}
\section{Findings}

\subsection{Outdated software components introduce avoidable vulnerabilities}
Focusing on three factors of update practice, excluding the way vendors deploy patches, our finding for all kinds of client hosts (including general purpose computing and IoT devices) is consistent with previous studies. Nappa et al. \cite{antonio-01} finds that devices move on to the latest version slowly showing a long tail, and Sarabi et al. \cite{sarabi-01} models how long they stay vulnerable (if they don"t update) using geometric distribution which also shows long tail. Our findings using the three metrics in the graph Figure \ref{fig:curl-vuln-version} and \ref{remaining-avoidable-plots} (Figure \ref{fig:okhttp-vuln-version}) for the four software over our capture duration suggest a similar trend, which tells us that our data is valid and devices are not lying about the user agents seen in our data.

In Figure \ref{fig:curl-vuln-version}, cURL is used as a user agent. On the horizontal axis we have a number of versions behind for different versions of user agents seen at different times during our data collection. The text label above the CDF for the version is the mean number of months older version is from the latest version available at the time of usage, for e.g., when a used version is 80 release versions behind, from the latest version it's on average 140 months (approximately 12 years) old. On the vertical axis on left, we have the distribution of the number of vulnerabilities that could have been avoided for every version behind, for e.g., when version behind is 25, number of avoidable vulnerabilities are in the range 21-35 (median 28) for critical and high severity vulnerabilities, and 24-39 (median 31.5) for vulnerabilities of all severities. On the vertical axis on right, we have a CDF plotted for the number of versions behind for 621 unique devices and cURL combinations. As per our data for cURL, there is a 50\% chance that when cURL is used as a user agent it's at least 14 release versions behind, 14 months old, and it could have avoided a median of 11 vulnerabilities by using the latest available version of that time. For cURL, at the long end of the tail we see 3 devices using more than 90 release ($\sim${}12 years) old versions. 

Plot for all the versions shows a similar long tail of older release version usage for all four software, see Figures \ref{fig:curl-vuln-version}, and \ref{remaining-avoidable-plots} (Figure \ref{fig:okhttp-vuln-version}). The number of avoidable vulnerabilities is different and depends upon how many vulnerabilities a software has. OkHtttp and Python Requests don't have many vulnerabilities, so updating to the latest version affects the vulnerability state to a lesser degree. From the plot of cURL and Wget, we can say that outdated versions over a period of time accrue more vulnerabilities and expand the host's vulnerability state. The previous statement is not true if a software doesn't have vulnerabilities, evident from the plot of Python Requests, and OkHttp.

We see surprising downtrend in number of avoidable vulnerabilities after number of version behind is more than 50 for cURL. cURL releases new version every 49 (median) days. So 50 and 90 version behind cURL is approximately 7 to 12 years old. Presumably the reason for the downtrend could be that cURL had less features 7 years ago, as its code base grew number of vulnerabilities increased.

\vspace{-10.0bp}
\subsection{Update practices of IoT devices are different from general computing devices}

To find the distinction in update practice (including vendor deployment and user installation) of IoT devices versus non-IoT devices we categorize devices in six categories as per Fingerbank's \cite{fingerbank} APIs. These six categories are: "IoT platforms", "IoT non-platforms" ,  "Computing", "Storage and Printers",  "Networking", and "Others", see Table \ref{device-cat}. 

\begin{table}[!ht]
    \centering
    \setlength{\abovecaptionskip}{2pt}
    \setlength{\belowcaptionskip}{-10pt}
    \caption{Device types in our 6 device categories based on Fingerbank API.}
    \begin{tabular}{p{0.30\linewidth} p{0.65\linewidth}}
    \toprule
    \textbf{Category} & \textbf{Device types} \\
    \midrule
    IoT platforms &  smart TVs, set top boxes (STB), and digital video recorders (DVR)\\
    IoT non-platforms & rest of the IoT devices, such as smart cameras, smart assistants, speakers, smart vacuum cleaners \\
    Computing & computers, phones, tablets, and Raspberry Pi kind of devices \\
    Storage and Printers &  network storage devices (NAS), and printer \\
    Networking & routers, wireless access points (WAP), switches, and firewalls\\
    Other & rest of the devices that can't be categorized by Fingerbank \\
    \bottomrule
    \end{tabular}
    \label{device-cat}
\end{table}

We distinguish between "IoT platforms" and "IoT non-platforms" as if an IoT device allows third-party applications to be installed on them, if it does we call them "IoT Platform". We distinguish "IoT platforms" from "IoT non-platforms" because applications installed on them could be affecting the version of user agents captured from those devices.  "Computing" category includes general purpose computing devices. "Storage and Printers" category contains network storage devices (NAS) and printers. "Networking" category contains networking devices found in homes. Other categories include devices that can't be classified by Fingerbank API into the previous five categories. Example of few vendors seen in our data are in Table \ref{software-vendors}.

\begin{table}[!ht]
    \centering
    \setlength{\abovecaptionskip}{2pt}
    \caption{Examples of few vendors we have seen for four software in our data.}
    \begin{tabular}{p{0.20\linewidth} p{0.75\linewidth}}
    \toprule
    \textbf{Software component} & \textbf{Few example vendors} \\ %
    \midrule
    cURL & Samsung SmartThings, Google Nest, Parrot IoT, Vizio TV, WD TV Live, Meraki WAP, Synology NAS, Roomba, etc. \\
    OkHttp & Amazon Alexa and Fire TV, Google Chromecast and Home, Roomba, Vizio TV, NVidia Shield, Belkin Router, etc. \\
    Wget & WD TV Live, Netgear, Synology NAS, QNAP NAS, Amazon Alexa, Tablo DVR, etc. \\
    Python Requests & Eero WAP, Cisco WAP, Synology NAS, TP-Link, NVidia Shield, Telldus Smart home, etc. \\
    \bottomrule
    \end{tabular}
    \label{software-vendors}
    \vspace{-1.0em}
\end{table}

\begin{figure*}[htbp]
    \centering
    \begin{subfigure}{1.0\linewidth}
        \centering
        \includegraphics[trim={0 1.6cm 0 2.0cm},clip, scale=0.50]{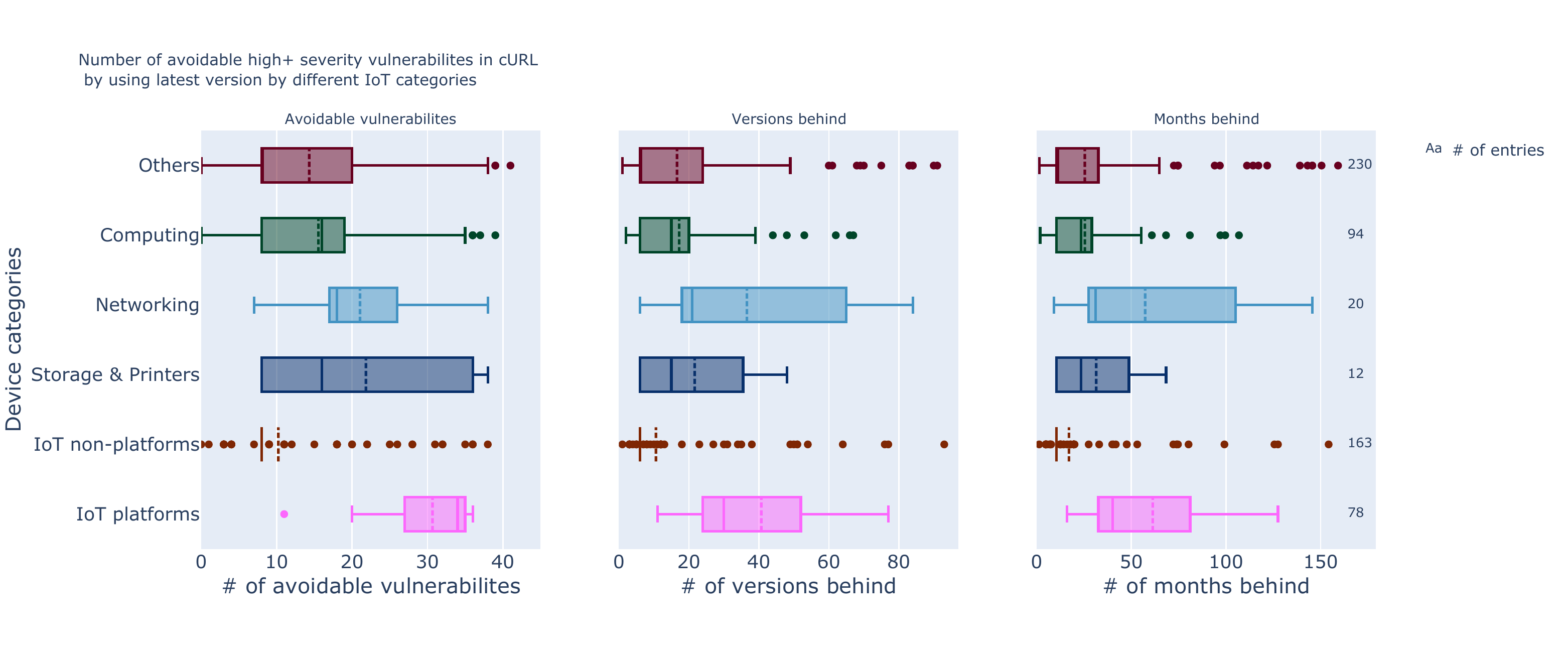}
    \end{subfigure}
    \begin{subfigure}{1.0\linewidth}
        \centering
        \includegraphics[trim={0 1cm 0 2.6cm},clip, scale=0.50]{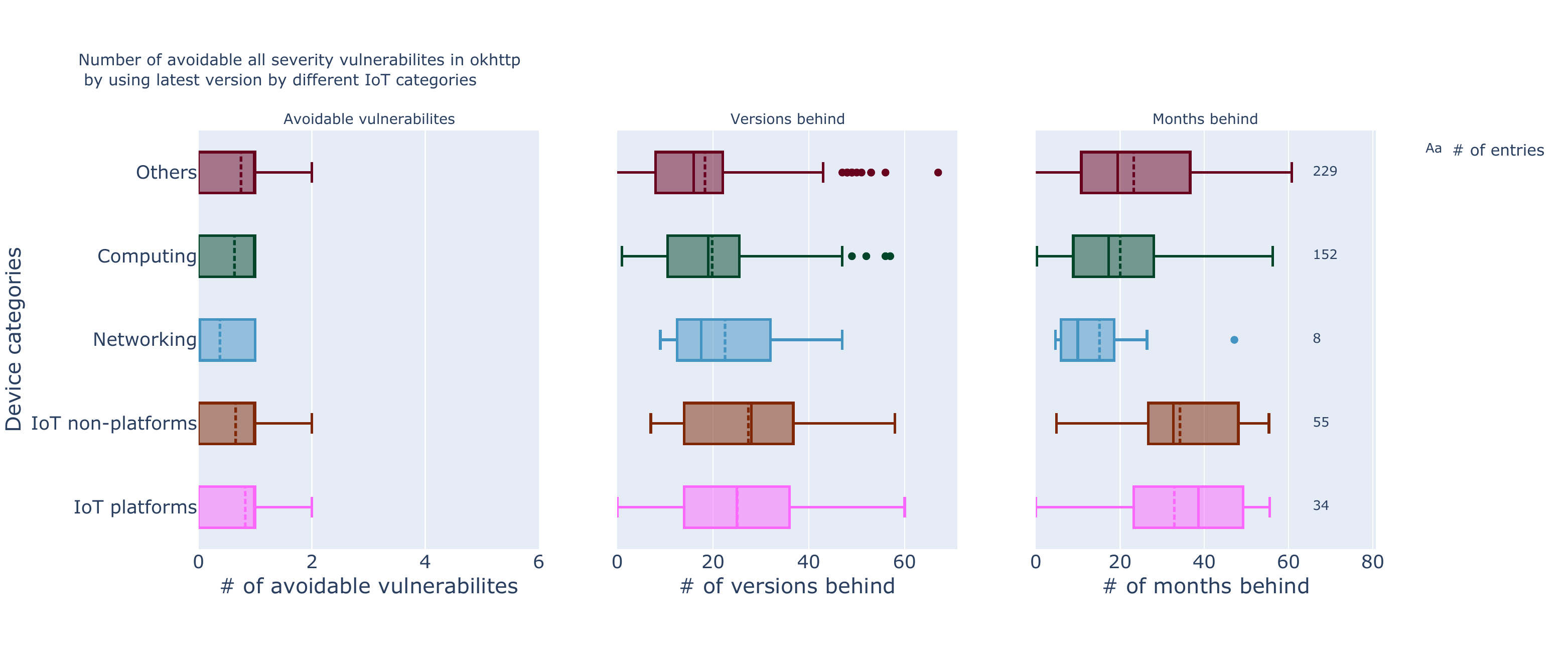}
    \end{subfigure}
    \setlength{\abovecaptionskip}{2pt}
    \caption{Distribution of number of avoidable vulnerabilities, number of versions behind, and number of months behind for all categories of IoT devices using cURL (top), and OkHttp (bottom).}
    \label{fig:curl-cat}
\end{figure*}

We box plot the distribution of vulnerabilities avoided (affects the vulnerability state of a device), number of versions behind, and number of days behind for each user agent seen in the earlier mentioned six categories for all the four software in Figures \ref{fig:curl-cat}, and Appendix \ref{remaining-distro-plots}. We would like to note that for OkHttp, Figure 4 shows that older versions (60 months old) and newer versions (20 - 25 months old) are equally vulnerable. This is because OkHttp has only two vulnerabilities and they exist in versions from 2017, and 2019. So either used versions of OkHTTP are five years old (after 2014), or two years old (after 2017) — they both contain two vulnerabilities and are equally vulnerable.

\subsubsection{Comparing "Computing" group with IoT devices}

From Figures \ref{fig:curl-cat}, and Appendix \ref{remaining-distro-plots} (Figure \ref{fig:wget-cat}), we could see that each category has a different distribution than others. By looking at the mean and median number of avoidable vulnerabilities, we can clearly see that personal computing devices have better software update posture than both the IoT device categories for all the four software, except for cURL in "IoT non-platforms" category. We looked further into this case, and found that there are 105 unique Samsung SmartThings \cite{smartthings} devices present in our dataset out of 163 devices in the "IoT non-platforms" category. These SmartThings devices consistently use the same cURL version 7.60.0, and are seen in a small burst of time. At the time when they are seen they are only 6 versions behind from the latest available version of cURL. These devices single-handedly improve the update posture of "IoT non-platforms" devices seen in our data by reducing the mean and median number of avoidable vulnerabilities of "IoT non-platforms" category. Focusing on the update practice of SmartThings over our data collection duration, we find their update practice is not good — they have two kinds of devices and both of them use the same version of cURL for more than 20 months. We conclude that these devices might have released an update at the time when we started data collection, and their presence in large numbers skewed the results for "IoT non-platforms category". If we exclude Samsung SmartThings devices, in the case of cURL, computing devices' update practice is better than "IoT non-platforms" and "IoT platforms" combined together. See mean and median values of number of avoidable vulnerabilities with and without SmartThings devices for these categories for cURL  in Table \ref{computing-v-iot}.

\begin{table}[!ht]
    \centering
    \setlength{\abovecaptionskip}{2pt}
    \setlength{\belowcaptionskip}{-10pt}
    \caption{Number of avoidable vulnerabilities in "Computing", "IoT platforms", and "IoT non-platforms" groups of cURL.}
    \begin{tabular}{p{0.50\linewidth} p{0.2\linewidth} p{0.2\linewidth}}
    \toprule
        \textbf{Group} & \textbf{mean} & \textbf{median} \\ 
        \midrule
        Computing & 15.54 & 16.0 \\ 
        IoT platforms & 30.66 & 34.0 \\ 
        IoT non-platforms & 10.22 & 8.0 \\ 
        IoT non-platforms without Samsung SmartThings & 15.18 & 13.5 \\
        Both IoT groups together & 16.83 & 8.0 \\
        Both groups together without Samsung SmartThings & 24.76 & 28.0 \\
        \bottomrule
    \end{tabular}
    \label{computing-v-iot}
    \vspace{-10bp}
\end{table}

\subsubsection{Comparing "IoT platforms" group with "IoT non-platforms"}

Focusing on IoT devices only, both the categories perform similarly in terms of updates practice if we take a look at the mean and median values of avoidable vulnerabilities for two IoT categories for all four software in the Appendix \ref{cat-stats-appendix} (Table \ref{table:cat-stats}), except for cURL for the reason explained in previous subsection. "IoT non-platforms" are doing better in general, which could be probably because large amount of third-party applications on platforms don't have as consistent update practice as a single vendor IoT devices.

\paragraph{Takeaway:} as previous studies have only focused on general purpose computing devices, we find that update practice of IoT devices are worse than computing devices.

\subsection{Possible slow vendor deployment and end user installation in IoT devices}

In general purpose computing, prior work \cite{thomas-01} has suggested that silent updates lead to quicker update installation compared to non-silent updates, and vendor update deployments depend upon vulnerability disclosure \cite{antonio-01}. For the software we used in our study, vulnerability disclosure factor for update practices is the same for all categories of devices, but the IoT group is using older versions for longer duration. That means the remaining three factors — user promptness, vendor deployment, and mechanism of deployment — are contributing to the reason why update practice of IoT devices is worse than general purpose computing devices. As these software are unlikely to be directly used by end users on IoT devices, updates for them go through the IoT vendors. Regarding vendor deployment factor in update, there has been an argument that vendors are bad for security, because they add an additional pit stop for the patches getting to end users \cite{ben-01, ben-02}. So maybe vendors are deploying patches slower than the computing group. User promptness to install patches is affected by update mechanism deployed by the vendors. Majority of well performing software used on devices in computing groups have now adopted the silent update strategy \cite{thomas-01}. Maybe IoT vendors are deploying patches promptly but not using silent update mechanisms, presumably for reasons like, not being able to do silent updates as IoT devices have constraints on computational power. In contrast with computing groups, some IoT devices are always on and don't need user engagement after the setup. To update IoT devices users have to use the companion apps, it could also be possible that users set up the IoT device and forget to check the companion app, which could be another reason for users' not promptly installing updates. Finding data relevant to all these factors could help us understand the actual reason behind sluggishness in IoT group update practices.

\begin{figure*}[!h]
    \centering
    \begin{subfigure}{1.0\linewidth}
        \centering
        \includegraphics[trim={0 0.5cm 0 0.5cm},clip,scale=0.50]{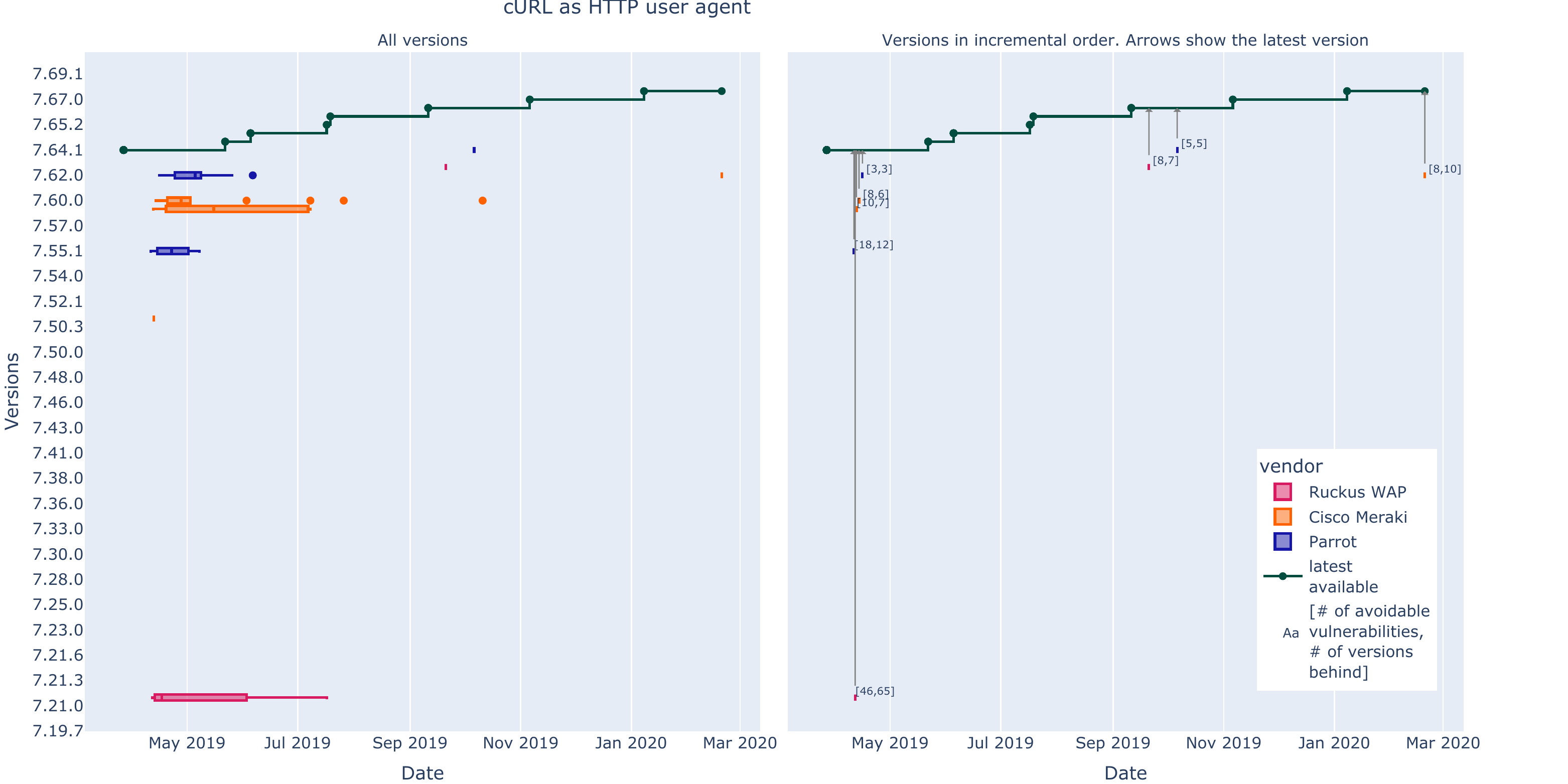}
    \end{subfigure}
    \begin{subfigure}{1.0\linewidth}
        \centering
        \includegraphics[trim={0 0 0 1.5cm},clip,scale=0.50]{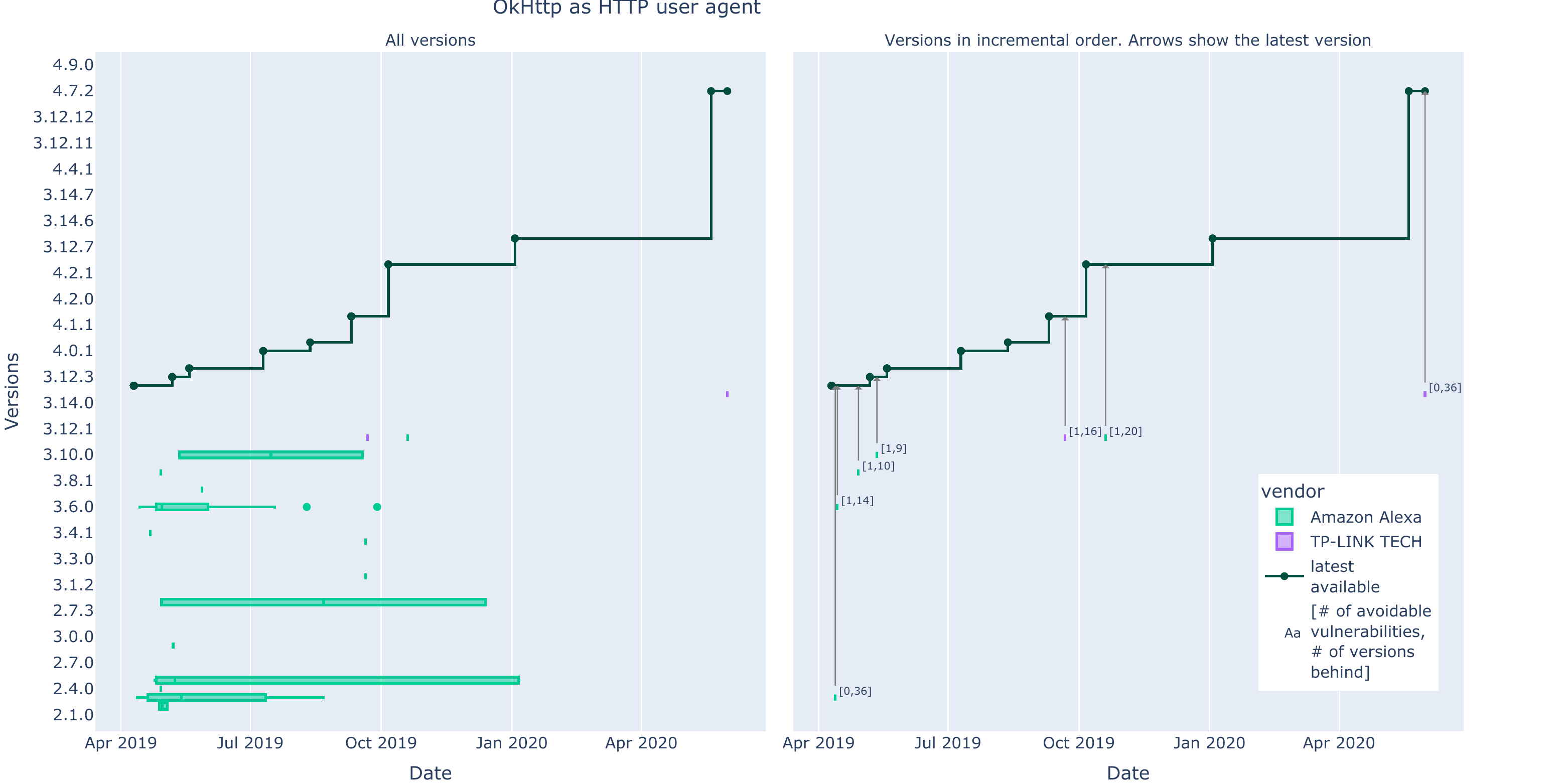}
    \end{subfigure}
    \setlength{\abovecaptionskip}{2pt}
    \setlength{\belowcaptionskip}{-10pt}
    \caption{For different vendors using cURL (top), and Wget (bottom), distribution of versions used against time in left subplot; version seen in incremental order with respect to time, along with number of avoidable vulnerabilities and number of version behind in square brackets in right subplot. Both the subplots show latest available version at the time of version usage.}
    \label{fig:curl-ttr}
\end{figure*}

\subsection{IoT vendors not deploying latest versions}
To understand the vendor deployment factor of update practice we plotted versions of four software used by user agent of different devices from different vendors over the duration of our data capture. We only used devices from "IoT non-platforms" for these plots because "IoT platforms" allow third party apps. Versions seen from "IoT platforms" category devices will not depict a single vendor's deployment practice. For both the vertical subplots in the Figure \ref{fig:curl-ttr}, and \ref{remaining-ttr} (Figure \ref{fig:req-ttr}) on the vertical axis we have different versions of software used, and time on the horizontal axis. All software is used by user agent of different vendors so we manually picked vendors which were seen more frequently in our data. Different colors of distribution represent different vendors as mentioned on the legends on the bottom right corner of every figure. Latest available version with respect to time is plotted in green color to give an understanding how far behind these used versions are from latest version. Plots also mention the metrics number of avoidable vulnerabilities and number of versions behind, described in section \ref{metric-explan}, in square brackets next to plotted distributions. 

From these previous plots we can see the trend that user agents of different devices of the same vendors are using the same version over a period of time. We also see that some devices from a vendor have updated to a newer version while some of them are still using older versions. This could possibly be explained by either IoT vendors are deploying updates in a rolling fashion or users are not installing the updates after the vendor deploy the updates. We do see a jump in versions after a period, which is shown on the right side subplot. We explain the right side plot in the next subsection \ref{ttr}.
\vspace{-1.0em}
\subsection{Time to repair (TTR)} \label{ttr}
We also try to calculate time to repair per version in terms of days for each software by calculating the number of days between two subsequent versions seen in our data and divide it by the number of release versions differences between them. A TTR is the slope between two points of the same vendor on the right side subplot.

Right subplot plots only include points when time and version both were seen in increasing order only. To elaborate, let's suppose that in April 2019, we see Alexa \cite{alexa} devices using version 2.4.0 of OkHttp on dates 1st, 10th, and 15. Again in 2019, we see older version 2.3.0 in May, a newer version 3.5.0 in July on dates 1st and 15th, and another newer version 2.5 on July 20th being used by Alexa devices. We include 2.4.0 used on 1st April and 3.5.0 used on 1st July in the right subplot because this is the first time we have seen these versions. Clearly suggesting that the vendor has deployed a patch as per our data. We don't include 2.3.0, used in May because the device has not been updated to a newer version, as we have seen newer versions from Alexa devices before on 1st April. We don't include 2.5.0 used on July 20th even though it's an increment in version from 2.4.0 because it might have been rolled out by the vendor earlier, sometime between 1st April and 1st July, it's just not present in our data, as we see 3.5.0 being used before on July 1st. 

\begin{table}[!h]
    \centering
    \setlength{\abovecaptionskip}{2pt}
    \setlength{\belowcaptionskip}{-5pt}
    \caption{Calculated TTR per version in days, hours:minutes:seconds for few vendors.}
    \begin{tabular}{p{0.30\linewidth} p{0.30\linewidth} p{0.30\linewidth}}
    \toprule
    \textbf{software} & \textbf{vendor}  & \textbf{TTR per version}\\ %
    \midrule
    cURL & Cisco Meraki & 1 days, 9:44:01\\
    & Cisco Meraki & 104 days, 0:30:41\\
    & Parrot & 0 days, 12:26:02\\
    & Parrot & 57 days, 20:46:07\\
    & Ruckus WAP & 2 days, 13:43:00\\
    Python requests & eero WAP & 0 days, 22:45:37\\
    & eero WAP & 320 days, 13:54:33\\
    OkHttp & Alexa & 0 days, 1:39:02\\
    & Alexa & 3 days, 16:50:06\\
    & Alexa & 6 days, 12:20:04\\
    & Alexa & 80 days, 9:25:09\\
    & TP-Link Tech & 63 days, 6:40:32\\
    \bottomrule
    \end{tabular}
    \label{ttr-table}
    \vspace{-1.0em}
\end{table}

Using this methodology we attempt to calculate the TTR per version of each vendor and report it in the Table \ref{ttr-table}. As our dataset is small, we have only few points in the right subplot in Figures \ref{fig:curl-ttr}. So TTRs we have calculated are few in terms of numbers. In the Table \ref{ttr-table}, the first small TTR for each version is because vendors probably rolled out the updates shortly after we started the capture, but the second ones are probably closer to actual TTR for the respective vendor. From this data we can say TTR for different vendors for the same software are different, consistent with \cite{antonio-01}.

\paragraph{Takeaways from our findings:}
\begin{itemize}

    \item IoT vendors are slower at deploying updates than all four software deploy updates, which could leave end users vulnerable if software has patched vulnerabilities in deployed updates.
    \item Even when IoT vendors do deploy updates they don't update to the latest version. 
    \item In case of cURL, the updated version could have avoided vulnerabilities by updating to the latest version available.
 
\end{itemize}

\section{Discussions and Future work}

Our dataset is small but good enough for a preliminary study of update practices in IoT ecosystem. Hopefully our preliminary findings are enough to point out to the research community that more effort is needed to gain understanding of update practice in the IoT ecosystem. We are planning to collect more data in future so that we can get more consistent, statistically large, and informative results, including (i) more device identification information (as we currently rely on FingerBank's API, which uses a blackbox machine learning model with unclear accuracy); and (ii) more evidence of software components beyond user agents (e.g., fingerprinting network traffic with p0f to infer the operating system~\cite{p0f}, or with TLS ClientHello to infer the SSL library and versions~\cite{TLSFingerprints}). 

Three factors out of four that are collectively responsible for update practice: user promptness, vendor update deployment, and mechanism of update deployment, we couldn't tell exactly which ones are contributing to worse update practice in IoT devices, compared to general purpose computing devices. Future studies could try to figure this out---for example, by asking IoT Inspector users about their actual update behaviors, thus collecting human-centered data alongside the User Agent data---so that efforts could be directed in the right direction to improve IoT ecosystem update practice.

We recommend IoT device users to install updates promptly upon notification. There are some IoT devices that don't require user interaction after installation, users should mindfully check for updates for these kinds of devices and install them promptly when available. We recommend vendors to use the latest versions of software components, even in cases when a component has few vulnerabilities (for e.g. OkHttp) because with one critical vulnerability attackers could do as much of damage as a software component having multiple critical vulnerabilities. We would also recommend vendors to keep SBOMs and track updates provided by software components they use. Learning from Linux vendors \cite{ben-01, ben-02, scott-01} , when software components provide updates, vendors should deploy those updates as soon as possible instead of acting as gatekeeper and blocking them because there is no other way for IoT device users to update those components. Vendors should deploy updates on IoT devices securely using standards like \cite{rfc4108, suit}, and silently (if possible) as previous studies have pointed out it’s the most effective way \cite{christos-01, stefan-01, thomas-01}.

\section{Summary}

We conduct a preliminary study of update practice of IoT devices and find that it's different from the findings of studies conducted about update practices of general purpose computing devices. We find that IoT devices are slower to update to newer versions compared to general purpose computing devices. We couldn't find the exact reason behind this but we narrowed down our findings to where future research should direct their efforts in order to find out which of three factors, end user promptness to install updates, vendor deployment practice, and vendor deployment mechanism are responsible. We also find that when IoT vendors deploy updates they don't update to the latest version, leaving hosts vulnerable to publicly known vulnerabilities.

\bibliographystyle{ACM-Reference-Format}
\bibliography{sample-base}


\begin{thebibliography}{52}


\ifx \showCODEN    \undefined \def \showCODEN     #1{\unskip}     \fi
\ifx \showDOI      \undefined \def \showDOI       #1{#1}\fi
\ifx \showISBNx    \undefined \def \showISBNx     #1{\unskip}     \fi
\ifx \showISBNxiii \undefined \def \showISBNxiii  #1{\unskip}     \fi
\ifx \showISSN     \undefined \def \showISSN      #1{\unskip}     \fi
\ifx \showLCCN     \undefined \def \showLCCN      #1{\unskip}     \fi
\ifx \shownote     \undefined \def \shownote      #1{#1}          \fi
\ifx \showarticletitle \undefined \def \showarticletitle #1{#1}   \fi
\ifx \showURL      \undefined \def \showURL       {\relax}        \fi
\providecommand\bibfield[2]{#2}
\providecommand\bibinfo[2]{#2}
\providecommand\natexlab[1]{#1}
\providecommand\showeprint[2][]{arXiv:#2}

\bibitem[IoT(2021)]%
        {IoTCompanies}
 \bibinfo{year}{2021}\natexlab{}.
\newblock \bibinfo{title}{The 1,200 IoT companies that are creating the
  connected world of the future -- IoT Startup Landscape 2021}.
\newblock
\newblock
\urldef\tempurl%
\url{https://iot-analytics.com/iot-startup-landscape/}
\showURL{%
\tempurl}


\bibitem[UA-(2022)]%
        {UA-Lying}
 \bibinfo{year}{2022}\natexlab{}.
\newblock \bibinfo{title}{Browser detection using the user agent}.
\newblock
\newblock
\urldef\tempurl%
\url{https://developer.mozilla.org/en-US/docs/Web/HTTP/Browser_detection_using_the_user_agent}
\showURL{%
\tempurl}


\bibitem[TLS(2022)]%
        {TLSFingerprints}
 \bibinfo{year}{2022}\natexlab{}.
\newblock \bibinfo{title}{Fingerprinting OpenSSL libraries on IoT devices}.
\newblock
\newblock
\urldef\tempurl%
\url{https://medium.com/all-things-inspected/fingerprinting-openssl-libraries-on-iot-devices-1d214b4d643}
\showURL{%
\tempurl}


\bibitem[p0f(2022)]%
        {p0f}
 \bibinfo{year}{2022}\natexlab{}.
\newblock \bibinfo{title}{p0f v3 (3.09b)}.
\newblock
\newblock
\urldef\tempurl%
\url{https://lcamtuf.coredump.cx/p0f3/}
\showURL{%
\tempurl}


\bibitem[sui(2022)]%
        {suit}
 \bibinfo{year}{2022}\natexlab{}.
\newblock \bibinfo{title}{Software Updates for Internet of Things (suit)}.
\newblock
\newblock
\urldef\tempurl%
\url{https://datatracker.ietf.org/wg/suit/about/}
\showURL{%
\tempurl}


\bibitem[Alhazmi and Malaiya(2005)]%
        {alhazmi-01}
\bibfield{author}{\bibinfo{person}{O.H. Alhazmi} {and} \bibinfo{person}{Y.K.
  Malaiya}.} \bibinfo{year}{2005}\natexlab{}.
\newblock \showarticletitle{Modeling the vulnerability discovery process}. In
  \bibinfo{booktitle}{\emph{16th IEEE International Symposium on Software
  Reliability Engineering (ISSRE'05)}}. \bibinfo{pages}{10 pp.--138}.
\newblock
\urldef\tempurl%
\url{https://doi.org/10.1109/ISSRE.2005.30}
\showDOI{\tempurl}


\bibitem[Alhazmi et~al\mbox{.}(2007)]%
        {alhazmi-02}
\bibfield{author}{\bibinfo{person}{O.H. Alhazmi}, \bibinfo{person}{Y.K.
  Malaiya}, {and} \bibinfo{person}{I. Ray}.} \bibinfo{year}{2007}\natexlab{}.
\newblock \showarticletitle{Measuring, analyzing and predicting security
  vulnerabilities in software systems}.
\newblock \bibinfo{journal}{\emph{Computers \& Security}} \bibinfo{volume}{26},
  \bibinfo{number}{3} (\bibinfo{year}{2007}), \bibinfo{pages}{219--228}.
\newblock
\showISSN{0167-4048}
\urldef\tempurl%
\url{https://doi.org/10.1016/j.cose.2006.10.002}
\showDOI{\tempurl}


\bibitem[Antonakakis et~al\mbox{.}(2017)]%
        {usenix_mirai}
\bibfield{author}{\bibinfo{person}{Manos Antonakakis}, \bibinfo{person}{Tim
  April}, \bibinfo{person}{Michael Bailey}, \bibinfo{person}{Matt Bernhard},
  \bibinfo{person}{Elie Bursztein}, \bibinfo{person}{Jaime Cochran},
  \bibinfo{person}{Zakir Durumeric}, \bibinfo{person}{J.~Alex Halderman},
  \bibinfo{person}{Luca Invernizzi}, \bibinfo{person}{Michalis Kallitsis},
  \bibinfo{person}{Deepak Kumar}, \bibinfo{person}{Chaz Lever},
  \bibinfo{person}{Zane Ma}, \bibinfo{person}{Joshua Mason},
  \bibinfo{person}{Damian Menscher}, \bibinfo{person}{Chad Seaman},
  \bibinfo{person}{Nick Sullivan}, \bibinfo{person}{Kurt Thomas}, {and}
  \bibinfo{person}{Yi Zhou}.} \bibinfo{year}{2017}\natexlab{}.
\newblock \showarticletitle{{Understanding the Mirai Botnet}}. In
  \bibinfo{booktitle}{\emph{USENIX Security Symposium}}.
\newblock


\bibitem[Arbaugh et~al\mbox{.}(2000)]%
        {arbaugh-01}
\bibfield{author}{\bibinfo{person}{W.A. Arbaugh}, \bibinfo{person}{W.L.
  Fithen}, {and} \bibinfo{person}{J. McHugh}.} \bibinfo{year}{2000}\natexlab{}.
\newblock \showarticletitle{Windows of vulnerability: a case study analysis}.
\newblock \bibinfo{journal}{\emph{Computer}} \bibinfo{volume}{33},
  \bibinfo{number}{12} (\bibinfo{year}{2000}), \bibinfo{pages}{52--59}.
\newblock
\urldef\tempurl%
\url{https://doi.org/10.1109/2.889093}
\showDOI{\tempurl}


\bibitem[Arora et~al\mbox{.}(2004)]%
        {arora-01}
\bibfield{author}{\bibinfo{person}{Ashish Arora}, \bibinfo{person}{Ramayya
  Krishnan}, \bibinfo{person}{Rahul Telang}, {and} \bibinfo{person}{Yubao
  Yang}.} \bibinfo{year}{2004}\natexlab{}.
\newblock \showarticletitle{Impact of Vulnerability Disclosure and Patch
  Availability - An Empirical Analysis}. In \bibinfo{booktitle}{\emph{In Third
  Workshop on the Economics of Information Security}}.
\newblock


\bibitem[Bettayeb et~al\mbox{.}(2019)]%
        {meriem-01}
\bibfield{author}{\bibinfo{person}{Meriem Bettayeb}, \bibinfo{person}{Qassim
  Nasir}, {and} \bibinfo{person}{Manar~Abu Talib}.}
  \bibinfo{year}{2019}\natexlab{}.
\newblock \showarticletitle{Firmware Update Attacks and Security for IoT
  Devices: Survey}. In \bibinfo{booktitle}{\emph{Proceedings of the ArabWIC 6th
  Annual International Conference Research Track}} (Rabat, Morocco)
  \emph{(\bibinfo{series}{ArabWIC 2019})}. \bibinfo{publisher}{Association for
  Computing Machinery}, \bibinfo{address}{New York, NY, USA}, Article
  \bibinfo{articleno}{4}, \bibinfo{numpages}{6}~pages.
\newblock
\showISBNx{9781450360890}
\urldef\tempurl%
\url{https://doi.org/10.1145/3333165.3333169}
\showDOI{\tempurl}


\bibitem[Bilge et~al\mbox{.}(2017)]%
        {bilge-01}
\bibfield{author}{\bibinfo{person}{Leyla Bilge}, \bibinfo{person}{Yufei Han},
  {and} \bibinfo{person}{Matteo Dell'Amico}.} \bibinfo{year}{2017}\natexlab{}.
\newblock \showarticletitle{RiskTeller: Predicting the Risk of Cyber
  Incidents}. In \bibinfo{booktitle}{\emph{Proceedings of the 2017 ACM SIGSAC
  Conference on Computer and Communications Security}} (Dallas, Texas, USA)
  \emph{(\bibinfo{series}{CCS '17})}. \bibinfo{publisher}{Association for
  Computing Machinery}, \bibinfo{address}{New York, NY, USA},
  \bibinfo{pages}{1299–1311}.
\newblock
\showISBNx{9781450349468}
\urldef\tempurl%
\url{https://doi.org/10.1145/3133956.3134022}
\showDOI{\tempurl}


\bibitem[Cavusoglu et~al\mbox{.}(2005)]%
        {cavusoglu-01}
\bibfield{author}{\bibinfo{person}{Hasan Cavusoglu}, \bibinfo{person}{Huseyin
  Cavusoglu}, {and} \bibinfo{person}{Srinivasan Raghunathan}.}
  \bibinfo{year}{2005}\natexlab{}.
\newblock \showarticletitle{Emerging Issues in Responsible Vulnerability
  Disclosure}. In \bibinfo{booktitle}{\emph{WEIS}}.
\newblock


\bibitem[Clark et~al\mbox{.}(2014)]%
        {sandy-01}
\bibfield{author}{\bibinfo{person}{Sandy Clark}, \bibinfo{person}{Michael
  Collis}, \bibinfo{person}{Matt Blaze}, {and} \bibinfo{person}{Jonathan~M.
  Smith}.} \bibinfo{year}{2014}\natexlab{}.
\newblock \showarticletitle{Moving Targets: Security and Rapid-Release in
  Firefox}. In \bibinfo{booktitle}{\emph{Proceedings of the 2014 ACM SIGSAC
  Conference on Computer and Communications Security}} (Scottsdale, Arizona,
  USA) \emph{(\bibinfo{series}{CCS '14})}. \bibinfo{publisher}{Association for
  Computing Machinery}, \bibinfo{address}{New York, NY, USA},
  \bibinfo{pages}{1256–1266}.
\newblock
\showISBNx{9781450329576}
\urldef\tempurl%
\url{https://doi.org/10.1145/2660267.2660320}
\showDOI{\tempurl}


\bibitem["cURL”(2022)]%
        {cURL}
\bibfield{author}{\bibinfo{person}{"cURL”}.} \bibinfo{year}{2022}\natexlab{}.
\newblock \bibinfo{booktitle}{\emph{"Command line tool and library for
  transferring data with URLs"}}.
\newblock
\urldef\tempurl%
\url{"https://curl.se/"}
\showURL{%
\tempurl}


\bibitem[DeKoven et~al\mbox{.}(2019)]%
        {louis-01}
\bibfield{author}{\bibinfo{person}{Louis~F. DeKoven}, \bibinfo{person}{Audrey
  Randall}, \bibinfo{person}{Ariana Mirian}, \bibinfo{person}{Gautam Akiwate},
  \bibinfo{person}{Ansel Blume}, \bibinfo{person}{Lawrence~K. Saul},
  \bibinfo{person}{Aaron Schulman}, \bibinfo{person}{Geoffrey~M. Voelker},
  {and} \bibinfo{person}{Stefan Savage}.} \bibinfo{year}{2019}\natexlab{}.
\newblock \showarticletitle{Measuring Security Practices and How They Impact
  Security}. In \bibinfo{booktitle}{\emph{Proceedings of the Internet
  Measurement Conference}} (Amsterdam, Netherlands) \emph{(\bibinfo{series}{IMC
  '19})}. \bibinfo{publisher}{Association for Computing Machinery},
  \bibinfo{address}{New York, NY, USA}, \bibinfo{pages}{36–49}.
\newblock
\showISBNx{9781450369480}
\urldef\tempurl%
\url{https://doi.org/10.1145/3355369.3355571}
\showDOI{\tempurl}


\bibitem[Duebendorfer and Frei(2009)]%
        {thomas-01}
\bibfield{author}{\bibinfo{person}{Thomas Duebendorfer} {and}
  \bibinfo{person}{Stefan Frei}.} \bibinfo{year}{2009}\natexlab{}.
\newblock \showarticletitle{Web Browser Security Update Effectiveness}. In
  \bibinfo{booktitle}{\emph{Proceedings of the 4th International Conference on
  Critical Information Infrastructures Security}} (Bonn, Germany)
  \emph{(\bibinfo{series}{CRITIS'09})}. \bibinfo{publisher}{Springer-Verlag},
  \bibinfo{address}{Berlin, Heidelberg}, \bibinfo{pages}{124–137}.
\newblock
\showISBNx{3642143784}


\bibitem[Durumeric et~al\mbox{.}(2014)]%
        {zakir-01}
\bibfield{author}{\bibinfo{person}{Zakir Durumeric}, \bibinfo{person}{Frank
  Li}, \bibinfo{person}{James Kasten}, \bibinfo{person}{Johanna Amann},
  \bibinfo{person}{Jethro Beekman}, \bibinfo{person}{Mathias Payer},
  \bibinfo{person}{Nicolas Weaver}, \bibinfo{person}{David Adrian},
  \bibinfo{person}{Vern Paxson}, \bibinfo{person}{Michael Bailey}, {and}
  \bibinfo{person}{J.~Alex Halderman}.} \bibinfo{year}{2014}\natexlab{}.
\newblock \showarticletitle{The Matter of Heartbleed}. In
  \bibinfo{booktitle}{\emph{Proceedings of the 2014 Conference on Internet
  Measurement Conference}} (Vancouver, BC, Canada) \emph{(\bibinfo{series}{IMC
  '14})}. \bibinfo{publisher}{Association for Computing Machinery},
  \bibinfo{address}{New York, NY, USA}, \bibinfo{pages}{475–488}.
\newblock
\showISBNx{9781450332132}
\urldef\tempurl%
\url{https://doi.org/10.1145/2663716.2663755}
\showDOI{\tempurl}


\bibitem["Fingerbank”(2022)]%
        {fingerbank}
\bibfield{author}{\bibinfo{person}{"Fingerbank”}.}
  \bibinfo{year}{2022}\natexlab{}.
\newblock \bibinfo{booktitle}{\emph{"ACCURATELY IDENTIFY CONNECTED DEVICES
  PERFORM ANOMALY DETECTION"}}.
\newblock
\urldef\tempurl%
\url{"https://www.fingerbank.org/about/"}
\showURL{%
\tempurl}


\bibitem[Frei et~al\mbox{.}(2009)]%
        {stefan-01}
\bibfield{author}{\bibinfo{person}{Stefan Frei}, \bibinfo{person}{Thomas
  Duebendorfer}, {and} \bibinfo{person}{Bernhard Plattner}.}
  \bibinfo{year}{2009}\natexlab{}.
\newblock \showarticletitle{Firefox (In) Security Update Dynamics Exposed}.
\newblock \bibinfo{journal}{\emph{SIGCOMM Comput. Commun. Rev.}}
  \bibinfo{volume}{39}, \bibinfo{number}{1} (\bibinfo{date}{dec}
  \bibinfo{year}{2009}), \bibinfo{pages}{16–22}.
\newblock
\showISSN{0146-4833}
\urldef\tempurl%
\url{https://doi.org/10.1145/1496091.1496094}
\showDOI{\tempurl}


\bibitem[Gkantsidis et~al\mbox{.}(2006)]%
        {christos-01}
\bibfield{author}{\bibinfo{person}{Christos Gkantsidis},
  \bibinfo{person}{Thomas Karagiannis}, {and} \bibinfo{person}{Milan
  VojnoviC}.} \bibinfo{year}{2006}\natexlab{}.
\newblock \showarticletitle{Planet Scale Software Updates}.
\newblock \bibinfo{journal}{\emph{SIGCOMM Comput. Commun. Rev.}}
  \bibinfo{volume}{36}, \bibinfo{number}{4} (\bibinfo{date}{aug}
  \bibinfo{year}{2006}), \bibinfo{pages}{423–434}.
\newblock
\showISSN{0146-4833}
\urldef\tempurl%
\url{https://doi.org/10.1145/1151659.1159961}
\showDOI{\tempurl}


\bibitem[Housley(2005)]%
        {rfc4108}
\bibfield{author}{\bibinfo{person}{Russ Housley}.}
  \bibinfo{year}{2005}\natexlab{}.
\newblock \bibinfo{title}{{Using Cryptographic Message Syntax (CMS) to Protect
  Firmware Packages}}.
\newblock \bibinfo{howpublished}{RFC 4108}.
\newblock
\urldef\tempurl%
\url{https://doi.org/10.17487/RFC4108}
\showDOI{\tempurl}


\bibitem[Huang et~al\mbox{.}(2020)]%
        {danny-01}
\bibfield{author}{\bibinfo{person}{Danny~Yuxing Huang}, \bibinfo{person}{Noah
  Apthorpe}, \bibinfo{person}{Frank Li}, \bibinfo{person}{Gunes Acar}, {and}
  \bibinfo{person}{Nick Feamster}.} \bibinfo{year}{2020}\natexlab{}.
\newblock \showarticletitle{IoT Inspector: Crowdsourcing Labeled Network
  Traffic from Smart Home Devices at Scale}.
\newblock \bibinfo{journal}{\emph{Proc. ACM Interact. Mob. Wearable Ubiquitous
  Technol.}} \bibinfo{volume}{4}, \bibinfo{number}{2}, Article
  \bibinfo{articleno}{46} (\bibinfo{date}{jun} \bibinfo{year}{2020}),
  \bibinfo{numpages}{21}~pages.
\newblock
\urldef\tempurl%
\url{https://doi.org/10.1145/3397333}
\showDOI{\tempurl}


\bibitem[Inc.(2022a)]%
        {alexa}
\bibfield{author}{\bibinfo{person}{Amazon.com Inc.}}
  \bibinfo{year}{2022}\natexlab{a}.
\newblock \bibinfo{booktitle}{\emph{Amazon Alexa}}.
\newblock
\urldef\tempurl%
\url{https://en.wikipedia.org/wiki/Amazon_Alexa}
\showURL{%
\tempurl}


\bibitem[Inc.(2022b)]%
        {smartthings}
\bibfield{author}{\bibinfo{person}{SmartThings Inc.}}
  \bibinfo{year}{2022}\natexlab{b}.
\newblock \bibinfo{booktitle}{\emph{Samsung SmartThings}}.
\newblock
\urldef\tempurl%
\url{https://en.wikipedia.org/wiki/SmartThings}
\showURL{%
\tempurl}


\bibitem[Kandek(2009)]%
        {vuln-law-2}
\bibfield{author}{\bibinfo{person}{Wolfgang Kandek}.}
  \bibinfo{year}{2009}\natexlab{}.
\newblock \bibinfo{booktitle}{\emph{The Laws of Vulnerabilities 2.0.}}
\newblock
\urldef\tempurl%
\url{https://www.qualys.com/docs/laws-of-vulnerabilities-2.0.pdf}
\showURL{%
\tempurl}


\bibitem[Khan et~al\mbox{.}(2012)]%
        {khan-01}
\bibfield{author}{\bibinfo{person}{Moazzam Khan}, \bibinfo{person}{Zehui Bi},
  {and} \bibinfo{person}{John~A. Copeland}.} \bibinfo{year}{2012}\natexlab{}.
\newblock \showarticletitle{Software updates as a security metric: Passive
  identification of update trends and effect on machine infection}. In
  \bibinfo{booktitle}{\emph{MILCOM 2012 - 2012 IEEE Military Communications
  Conference}}. \bibinfo{pages}{1--6}.
\newblock
\urldef\tempurl%
\url{https://doi.org/10.1109/MILCOM.2012.6415869}
\showDOI{\tempurl}


\bibitem[Kim et~al\mbox{.}(2018)]%
        {kim-01}
\bibfield{author}{\bibinfo{person}{Dae-Young Kim}, \bibinfo{person}{Seokhoon
  Kim}, {and} \bibinfo{person}{Jong~Hyuk Park}.}
  \bibinfo{year}{2018}\natexlab{}.
\newblock \showarticletitle{Remote Software Update in Trusted Connection of
  Long Range IoT Networking Integrated With Mobile Edge Cloud}.
\newblock \bibinfo{journal}{\emph{IEEE Access}}  \bibinfo{volume}{6}
  (\bibinfo{year}{2018}), \bibinfo{pages}{66831--66840}.
\newblock
\urldef\tempurl%
\url{https://doi.org/10.1109/ACCESS.2017.2774239}
\showDOI{\tempurl}


\bibitem[Kumar et~al\mbox{.}(2019)]%
        {kumar2019all}
\bibfield{author}{\bibinfo{person}{Deepak Kumar}, \bibinfo{person}{Kelly Shen},
  \bibinfo{person}{Benton Case}, \bibinfo{person}{Deepali Garg},
  \bibinfo{person}{Galina Alperovich}, \bibinfo{person}{Dmitry Kuznetsov},
  \bibinfo{person}{Rajarshi Gupta}, {and} \bibinfo{person}{Zakir Durumeric}.}
  \bibinfo{year}{2019}\natexlab{}.
\newblock \showarticletitle{All Things Considered: An Analysis of $\{$IoT$\}$
  Devices on Home Networks}. In \bibinfo{booktitle}{\emph{28th USENIX security
  symposium (USENIX Security 19)}}. \bibinfo{pages}{1169--1185}.
\newblock


\bibitem[Langiu et~al\mbox{.}(2019)]%
        {langiu-01}
\bibfield{author}{\bibinfo{person}{Antonio Langiu},
  \bibinfo{person}{Carlo~Alberto Boano}, \bibinfo{person}{Markus Schuß}, {and}
  \bibinfo{person}{Kay Römer}.} \bibinfo{year}{2019}\natexlab{}.
\newblock \showarticletitle{UpKit: An Open-Source, Portable, and Lightweight
  Update Framework for Constrained IoT Devices}. In
  \bibinfo{booktitle}{\emph{2019 IEEE 39th International Conference on
  Distributed Computing Systems (ICDCS)}}. \bibinfo{pages}{2101--2112}.
\newblock
\urldef\tempurl%
\url{https://doi.org/10.1109/ICDCS.2019.00207}
\showDOI{\tempurl}


\bibitem[Laurie(2008a)]%
        {ben-01}
\bibfield{author}{\bibinfo{person}{Ben Laurie}.}
  \bibinfo{year}{2008}\natexlab{a}.
\newblock \bibinfo{booktitle}{\emph{Debian and OpenSSL: The Aftermath}}.
\newblock
\urldef\tempurl%
\url{https://www.links.org/?p=328}
\showURL{%
\tempurl}


\bibitem[Laurie(2008b)]%
        {ben-02}
\bibfield{author}{\bibinfo{person}{Ben Laurie}.}
  \bibinfo{year}{2008}\natexlab{b}.
\newblock \bibinfo{booktitle}{\emph{Vendors Are Bad For Security}}.
\newblock
\urldef\tempurl%
\url{https://www.links.org/?p=327}
\showURL{%
\tempurl}


\bibitem[Mohajeri~Moghaddam et~al\mbox{.}(2019)]%
        {mohajeri2019watching}
\bibfield{author}{\bibinfo{person}{Hooman Mohajeri~Moghaddam},
  \bibinfo{person}{Gunes Acar}, \bibinfo{person}{Ben Burgess},
  \bibinfo{person}{Arunesh Mathur}, \bibinfo{person}{Danny~Yuxing Huang},
  \bibinfo{person}{Nick Feamster}, \bibinfo{person}{Edward~W Felten},
  \bibinfo{person}{Prateek Mittal}, {and} \bibinfo{person}{Arvind Narayanan}.}
  \bibinfo{year}{2019}\natexlab{}.
\newblock \showarticletitle{Watching you watch: The tracking ecosystem of
  over-the-top tv streaming devices}. In \bibinfo{booktitle}{\emph{Proceedings
  of the 2019 ACM SIGSAC Conference on Computer and Communications Security}}.
  \bibinfo{pages}{131--147}.
\newblock


\bibitem[Morgner et~al\mbox{.}(2020)]%
        {philipp-01}
\bibfield{author}{\bibinfo{person}{Philipp Morgner}, \bibinfo{person}{Christoph
  Mai}, \bibinfo{person}{Nicole Koschate-Fischer}, \bibinfo{person}{Felix
  Freiling}, {and} \bibinfo{person}{Zinaida Benenson}.}
  \bibinfo{year}{2020}\natexlab{}.
\newblock \showarticletitle{Security Update Labels: Establishing Economic
  Incentives for Security Patching of IoT Consumer Products}. In
  \bibinfo{booktitle}{\emph{2020 IEEE Symposium on Security and Privacy (SP)}}.
  \bibinfo{pages}{429--446}.
\newblock
\urldef\tempurl%
\url{https://doi.org/10.1109/SP40000.2020.00021}
\showDOI{\tempurl}


\bibitem[Nappa et~al\mbox{.}(2015)]%
        {antonio-01}
\bibfield{author}{\bibinfo{person}{Antonio Nappa}, \bibinfo{person}{Richard
  Johnson}, \bibinfo{person}{Leyla Bilge}, \bibinfo{person}{Juan Caballero},
  {and} \bibinfo{person}{Tudor Dumitras}.} \bibinfo{year}{2015}\natexlab{}.
\newblock \showarticletitle{The Attack of the Clones: A Study of the Impact of
  Shared Code on Vulnerability Patching}. In \bibinfo{booktitle}{\emph{2015
  IEEE Symposium on Security and Privacy}}. \bibinfo{pages}{692--708}.
\newblock
\urldef\tempurl%
\url{https://doi.org/10.1109/SP.2015.48}
\showDOI{\tempurl}


\bibitem[Neuhaus et~al\mbox{.}(2007)]%
        {neuhaus-01}
\bibfield{author}{\bibinfo{person}{Stephan Neuhaus}, \bibinfo{person}{Thomas
  Zimmermann}, \bibinfo{person}{Christian Holler}, {and}
  \bibinfo{person}{Andreas Zeller}.} \bibinfo{year}{2007}\natexlab{}.
\newblock \showarticletitle{Predicting Vulnerable Software Components}. In
  \bibinfo{booktitle}{\emph{Proceedings of the 14th ACM Conference on Computer
  and Communications Security}} (Alexandria, Virginia, USA)
  \emph{(\bibinfo{series}{CCS '07})}. \bibinfo{publisher}{Association for
  Computing Machinery}, \bibinfo{address}{New York, NY, USA},
  \bibinfo{pages}{529–540}.
\newblock
\showISBNx{9781595937032}
\urldef\tempurl%
\url{https://doi.org/10.1145/1315245.1315311}
\showDOI{\tempurl}


\bibitem[{NVD}(2022)]%
        {nvd}
\bibfield{author}{\bibinfo{person}{{NVD}}.} \bibinfo{year}{2022}\natexlab{}.
\newblock \bibinfo{booktitle}{\emph{National Vulnerability Database}}.
\newblock NIST.
\newblock
\urldef\tempurl%
\url{https://nvd.nist.gov/}
\showURL{%
\tempurl}


\bibitem["OkHttp”(2022)]%
        {okhttp}
\bibfield{author}{\bibinfo{person}{"OkHttp”}.}
  \bibinfo{year}{2022}\natexlab{}.
\newblock \bibinfo{booktitle}{\emph{"OkHttp HTTP client"}}.
\newblock
\urldef\tempurl%
\url{"https://square.github.io/okhttp/"}
\showURL{%
\tempurl}


\bibitem[Paracha et~al\mbox{.}(2021)]%
        {paracha2021iotls}
\bibfield{author}{\bibinfo{person}{Muhammad~Talha Paracha},
  \bibinfo{person}{Daniel~J Dubois}, \bibinfo{person}{Narseo
  Vallina-Rodriguez}, {and} \bibinfo{person}{David Choffnes}.}
  \bibinfo{year}{2021}\natexlab{}.
\newblock \showarticletitle{IoTLS: understanding TLS usage in consumer IoT
  devices}. In \bibinfo{booktitle}{\emph{Proceedings of the 21st ACM Internet
  Measurement Conference}}. \bibinfo{pages}{165--178}.
\newblock


\bibitem[Ramos(2006)]%
        {vuln-law-1}
\bibfield{author}{\bibinfo{person}{Terry Ramos}.}
  \bibinfo{year}{2006}\natexlab{}.
\newblock \bibinfo{booktitle}{\emph{The Laws of Vulnerabilities}}.
\newblock
\urldef\tempurl%
\url{https://www.qualys.com/docs/laws-of-vulnerabilities-presentation.pdf}
\showURL{%
\tempurl}


\bibitem[Reeder et~al\mbox{.}(2017)]%
        {robert-01}
\bibfield{author}{\bibinfo{person}{Robert~W. Reeder}, \bibinfo{person}{Iulia
  Ion}, {and} \bibinfo{person}{Sunny Consolvo}.}
  \bibinfo{year}{2017}\natexlab{}.
\newblock \showarticletitle{152 Simple Steps to Stay Safe Online: Security
  Advice for Non-Tech-Savvy Users}.
\newblock \bibinfo{journal}{\emph{IEEE Security and Privacy}}
  (\bibinfo{year}{2017}).
\newblock


\bibitem[Ren et~al\mbox{.}(2019)]%
        {ren2019information}
\bibfield{author}{\bibinfo{person}{Jingjing Ren}, \bibinfo{person}{Daniel~J.
  Dubois}, \bibinfo{person}{David~R. Choffnes}, \bibinfo{person}{Anna~Maria
  Mandalari}, \bibinfo{person}{Roman Kolcun}, {and} \bibinfo{person}{Hamed
  Haddadi}.} \bibinfo{year}{2019}\natexlab{}.
\newblock \showarticletitle{Information Exposure From Consumer IoT Devices: {A}
  Multidimensional, Network-Informed Measurement Approach}. In
  \bibinfo{booktitle}{\emph{Proceedings of the Internet Measurement
  Conference}}. \bibinfo{pages}{267--279}.
\newblock


\bibitem[Requests”(2022)]%
        {requests}
\bibfield{author}{\bibinfo{person}{"Python Requests”}.}
  \bibinfo{year}{2022}\natexlab{}.
\newblock \bibinfo{booktitle}{\emph{"an elegant and simple HTTP library for
  Python, built for human beings"}}.
\newblock
\urldef\tempurl%
\url{"https://en.wikipedia.org/wiki/Requests_(software)"}
\showURL{%
\tempurl}


\bibitem[Rescorla(2003)]%
        {eric-01}
\bibfield{author}{\bibinfo{person}{Eric Rescorla}.}
  \bibinfo{year}{2003}\natexlab{}.
\newblock \showarticletitle{Security Holes . . . Who Cares?}. In
  \bibinfo{booktitle}{\emph{12th USENIX Security Symposium (USENIX Security
  03)}}. \bibinfo{publisher}{USENIX Association}, \bibinfo{address}{Washington,
  D.C.}
\newblock
\urldef\tempurl%
\url{https://www.usenix.org/conference/12th-usenix-security-symposium/security-holes-who-cares}
\showURL{%
\tempurl}


\bibitem[Rescorla(2005)]%
        {eric-02}
\bibfield{author}{\bibinfo{person}{Eric Rescorla}.}
  \bibinfo{year}{2005}\natexlab{}.
\newblock \showarticletitle{Is Finding Security Holes a Good Idea?}
\newblock \bibinfo{journal}{\emph{IEEE Security and Privacy}}
  \bibinfo{volume}{3}, \bibinfo{number}{1} (\bibinfo{date}{jan}
  \bibinfo{year}{2005}), \bibinfo{pages}{14–19}.
\newblock
\showISSN{1540-7993}
\urldef\tempurl%
\url{https://doi.org/10.1109/MSP.2005.17}
\showDOI{\tempurl}


\bibitem[Sarabi et~al\mbox{.}(2017)]%
        {sarabi-01}
\bibfield{author}{\bibinfo{person}{Armin Sarabi}, \bibinfo{person}{Ziyun Zhu},
  \bibinfo{person}{Chaowei Xiao}, \bibinfo{person}{Mingyan Liu}, {and}
  \bibinfo{person}{Tudor Dumitras}.} \bibinfo{year}{2017}\natexlab{}.
\newblock \showarticletitle{Patch Me If You Can: A Study on the Effects of
  Individual User Behavior on the End-Host Vulnerability State}. In
  \bibinfo{booktitle}{\emph{Proceedings of the 18th Passive and Active
  Measurement PAM.}} \bibinfo{publisher}{ACM}, \bibinfo{address}{Sydney,
  Australia.}
\newblock


\bibitem[Shahzad et~al\mbox{.}(2012)]%
        {shahzad-01}
\bibfield{author}{\bibinfo{person}{Muhammad Shahzad},
  \bibinfo{person}{Muhammad~Zubair Shafiq}, {and} \bibinfo{person}{Alex~X.
  Liu}.} \bibinfo{year}{2012}\natexlab{}.
\newblock \showarticletitle{A large scale exploratory analysis of software
  vulnerability life cycles}. In \bibinfo{booktitle}{\emph{2012 34th
  International Conference on Software Engineering (ICSE)}}.
  \bibinfo{pages}{771--781}.
\newblock
\urldef\tempurl%
\url{https://doi.org/10.1109/ICSE.2012.6227141}
\showDOI{\tempurl}


\bibitem["Wget”(2022)]%
        {Wget}
\bibfield{author}{\bibinfo{person}{"Wget”}.} \bibinfo{year}{2022}\natexlab{}.
\newblock \bibinfo{booktitle}{\emph{"Retrieve files via HTTP or FTP"}}.
\newblock
\urldef\tempurl%
\url{"https://en.wikipedia.org/wiki/Wget"}
\showURL{%
\tempurl}


\bibitem[Yilek et~al\mbox{.}(2009)]%
        {scott-01}
\bibfield{author}{\bibinfo{person}{Scott Yilek}, \bibinfo{person}{Eric
  Rescorla}, \bibinfo{person}{Hovav Shacham}, \bibinfo{person}{Brandon
  Enright}, {and} \bibinfo{person}{Stefan Savage}.}
  \bibinfo{year}{2009}\natexlab{}.
\newblock \showarticletitle{When Private Keys Are Public: Results from the 2008
  Debian OpenSSL Vulnerability}. In \bibinfo{booktitle}{\emph{Proceedings of
  the 9th ACM SIGCOMM Conference on Internet Measurement}} (Chicago, Illinois,
  USA) \emph{(\bibinfo{series}{IMC '09})}. \bibinfo{publisher}{Association for
  Computing Machinery}, \bibinfo{address}{New York, NY, USA},
  \bibinfo{pages}{15–27}.
\newblock
\showISBNx{9781605587714}
\urldef\tempurl%
\url{https://doi.org/10.1145/1644893.1644896}
\showDOI{\tempurl}


\bibitem[Yohan and Lo(2018)]%
        {alexander-01}
\bibfield{author}{\bibinfo{person}{Alexander Yohan} {and}
  \bibinfo{person}{Nai-Wei Lo}.} \bibinfo{year}{2018}\natexlab{}.
\newblock \showarticletitle{An Over-the-Blockchain Firmware Update Framework
  for IoT Devices}. In \bibinfo{booktitle}{\emph{2018 IEEE Conference on
  Dependable and Secure Computing (DSC)}}. \bibinfo{pages}{1--8}.
\newblock
\urldef\tempurl%
\url{https://doi.org/10.1109/DESEC.2018.8625164}
\showDOI{\tempurl}


\bibitem[Yousefnezhad et~al\mbox{.}(2020)]%
        {narges-01}
\bibfield{author}{\bibinfo{person}{Narges Yousefnezhad},
  \bibinfo{person}{Avleen Malhi}, {and} \bibinfo{person}{Kary Främling}.}
  \bibinfo{year}{2020}\natexlab{}.
\newblock \showarticletitle{Security in product lifecycle of IoT devices: A
  survey}.
\newblock \bibinfo{journal}{\emph{Journal of Network and Computer
  Applications}}  \bibinfo{volume}{171} (\bibinfo{year}{2020}),
  \bibinfo{pages}{102779}.
\newblock
\showISSN{1084-8045}
\urldef\tempurl%
\url{https://doi.org/10.1016/j.jnca.2020.102779}
\showDOI{\tempurl}


\bibitem[Zhang et~al\mbox{.}(2016)]%
        {chi-01}
\bibfield{author}{\bibinfo{person}{Chi Zhang}, \bibinfo{person}{Wonsun Ahn},
  \bibinfo{person}{Youtao Zhang}, {and} \bibinfo{person}{Bruce~R. Childers}.}
  \bibinfo{year}{2016}\natexlab{}.
\newblock \showarticletitle{Live code update for IoT devices in energy
  harvesting environments}. In \bibinfo{booktitle}{\emph{2016 5th Non-Volatile
  Memory Systems and Applications Symposium (NVMSA)}}. \bibinfo{pages}{1--6}.
\newblock
\urldef\tempurl%
\url{https://doi.org/10.1109/NVMSA.2016.7547182}
\showDOI{\tempurl}


\end{thebibliography}

\appendix

\begin{figure*}[!t]
    \centering
    \setlength{\abovecaptionskip}{2pt}
    \setlength{\belowcaptionskip}{-5pt}
    \begin{subfigure}[c]{1.0\linewidth}
        \centering
        \includegraphics[trim={0 1.2cm 0 2.6cm},clip,scale=0.40]{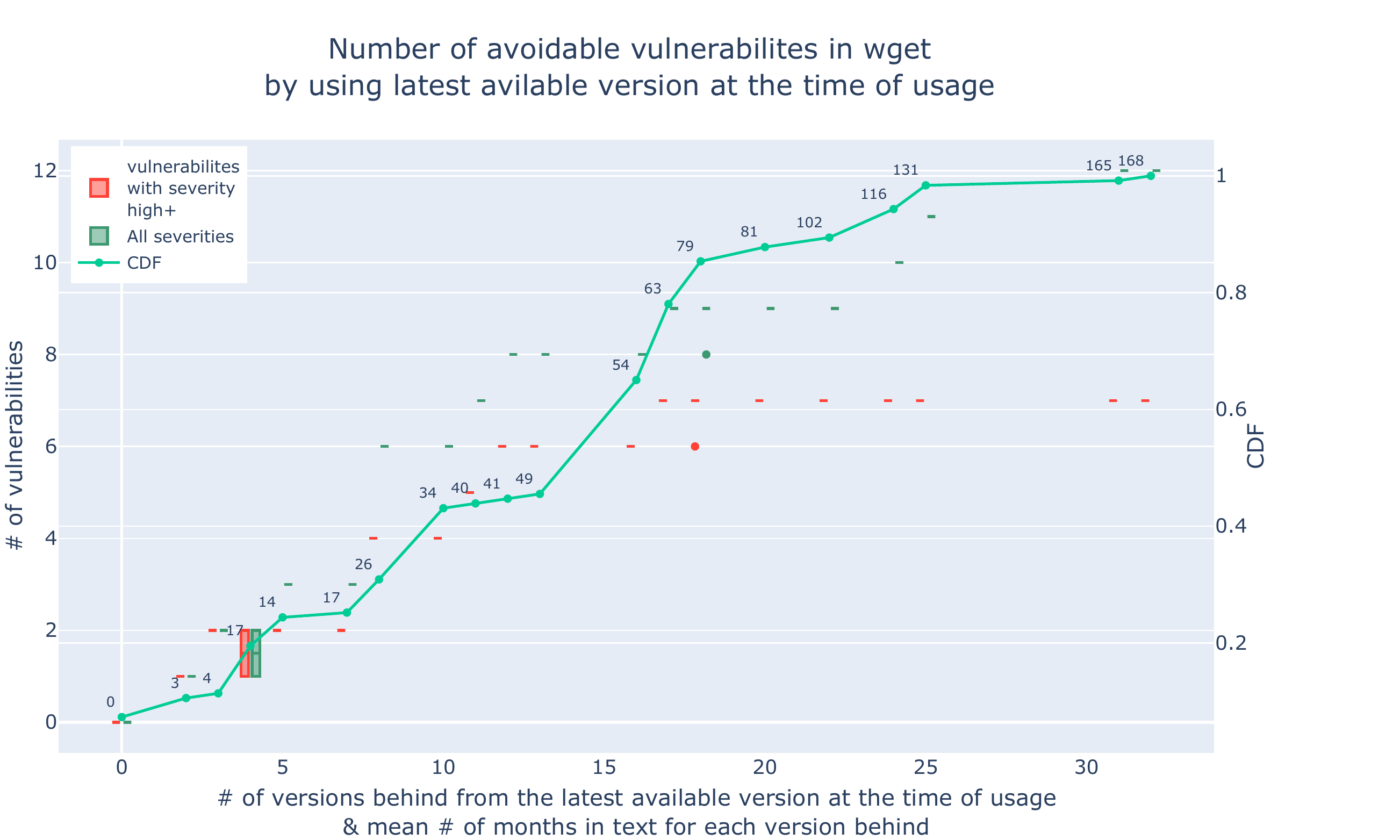}
    \end{subfigure}
    \begin{subfigure}{1.0\linewidth}
        \centering
        \includegraphics[trim={0 1.2cm 0 2.2cm},clip,scale=0.45]{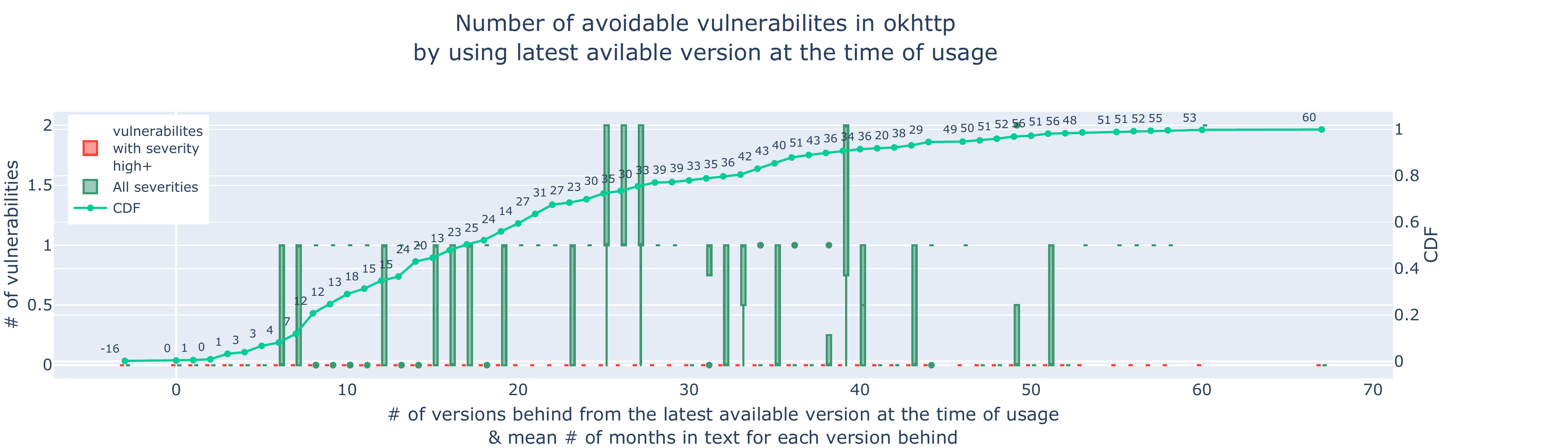}
    \end{subfigure}
    \begin{subfigure}{1.0\linewidth}
        \centering
        \includegraphics[trim={0 0 0 2.2cm},clip,scale=0.45]{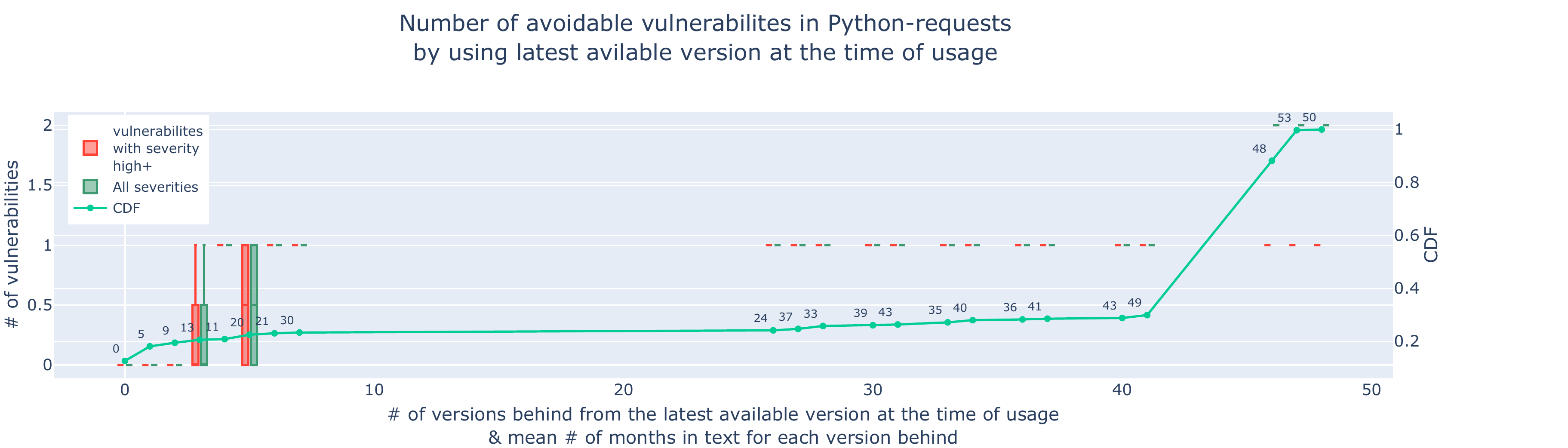}
    \end{subfigure}
    \caption{Number of avoidable vulnerabilities in Wget, OkHttp, and Python Requests, from top to bottom, against number of versions behind, and CDF distribution of devices for number of versions behind.}
    \label{fig:okhttp-vuln-version}
\end{figure*}

\begin{figure*}[!h]
    \centering
    \setlength{\abovecaptionskip}{2pt}
    \setlength{\belowcaptionskip}{-5pt}
    \begin{subfigure}{1.0\linewidth}
        \centering
        \includegraphics[trim={0 1.6cm 0.0cm 2.0cm},clip, scale=0.42]{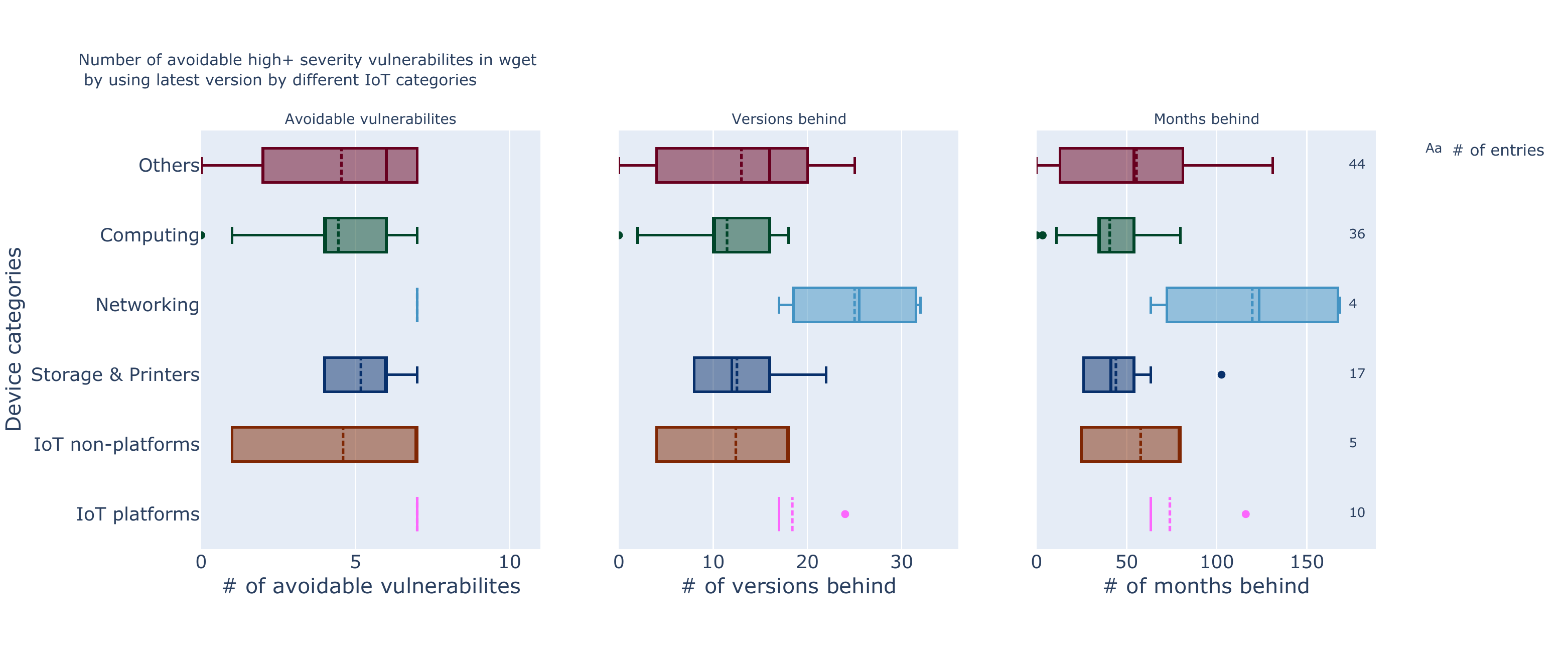}
    \end{subfigure}
    \begin{subfigure}{1.0\linewidth}
        \centering
        \includegraphics[trim={0 1cm 0cm 2.6cm},clip, scale=0.42]{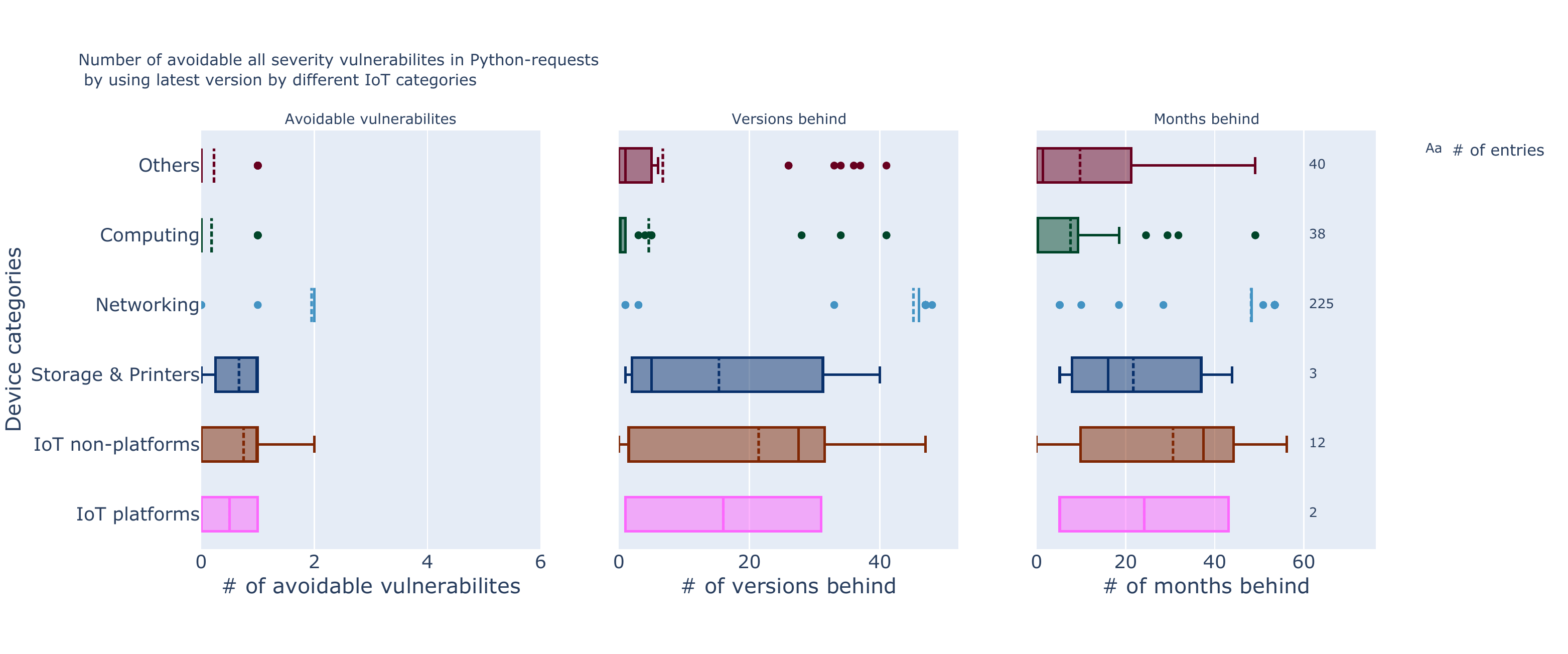}
    \end{subfigure}
    \caption{Distribution of number of avoidable vulnerabilities, number of versions behind, and number of months behind for all categories of IoT devices using Wget (top), and Python Requests (bottom).}
    \label{fig:wget-cat}
\end{figure*}

\begin{figure*}[!t]
    \centering
    \setlength{\abovecaptionskip}{2pt}
    \includegraphics[trim={0 0 0 0.5cm},clip,scale=0.6]{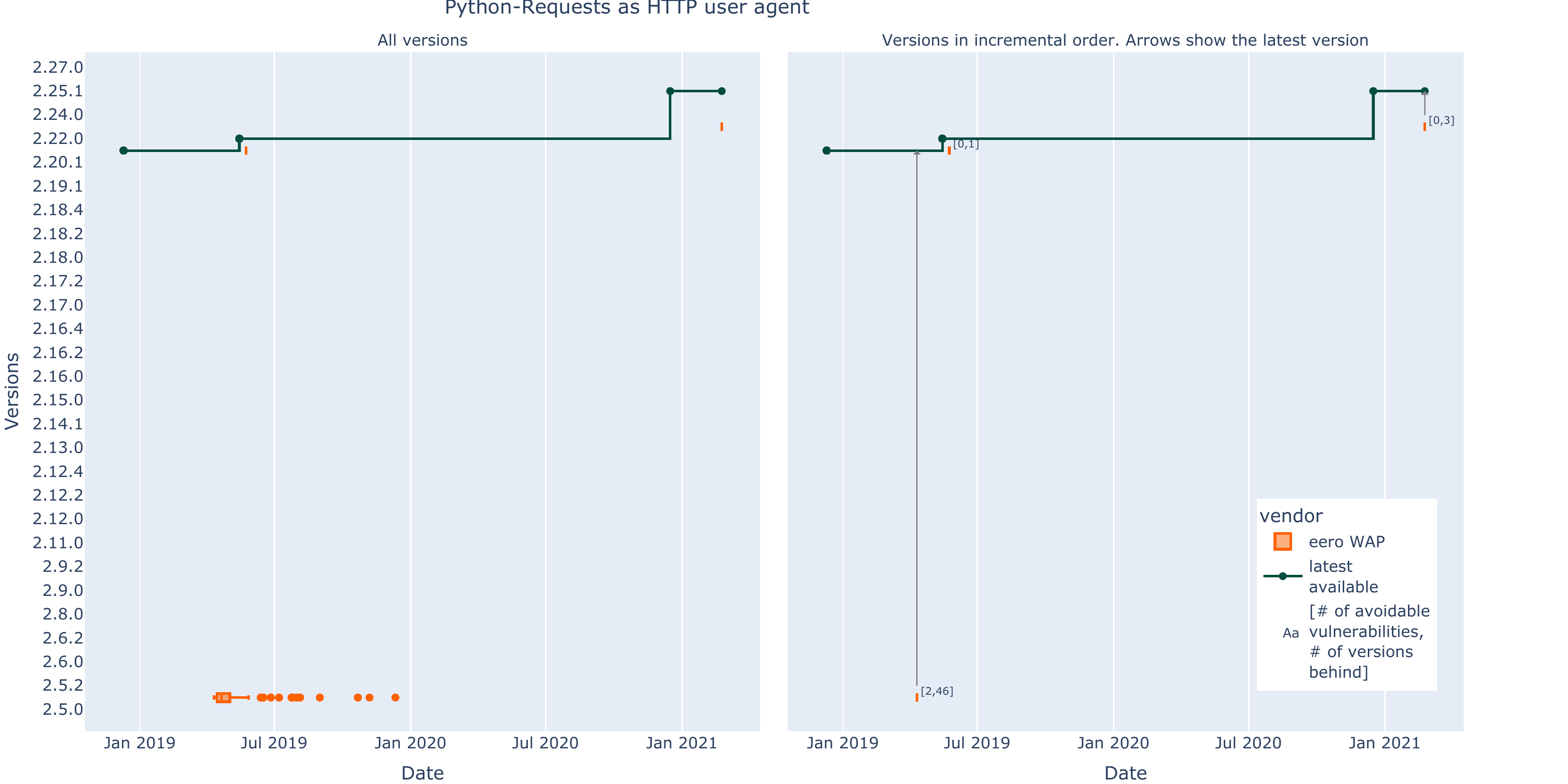}
    \caption{For different vendors using Python Requests, distribution of versions used against time in left subplot; version seen in incremental order with respect to time, along with number of avoidable vulnerabilities and number of version behind in square brackets in right subplot. Both the subplots show latest available version at the time of version usage.}
    \label{fig:req-ttr}
\end{figure*}

\begin{table*}[t]
    \centering
    \setlength{\abovecaptionskip}{2pt}
    \caption{Table of mean, and median (in parenthesis) for each categories of IoT devices' distribution of number of avoidable vulnerabilities, number of versions behind, and number of months behind for all software.}
    \begin{tabular}{p{0.05\linewidth} p{0.20\linewidth} p{0.11 \linewidth} p{0.11\linewidth}
        p{0.11\linewidth} p{0.11\linewidth} p{0.11\linewidth} p{0.11\linewidth}}       
    \toprule
        \textbf{prog} & \textbf{metric} & \textbf{IoT platforms} & \textbf{IoT non-platforms} & \textbf{Storage \& Printers} & \textbf{Networking} & \textbf{Computing} & \textbf{Others} \\ 
    \midrule
        cURL & avoidable hi+ vulnerabilities & 30.67 (34.00) & 10.22 (8.00) & 21.83 (16.00) & 21.05 (18.00) & 15.54 (16.00) & 14.31 (8.00) \\
         & versions behind & 40.71 (30.00) & 10.58 (6.00) & 21.67 (15.00) & 36.60 (21.00) & 17.23 (15.00) & 16.60 (6.00) \\
         & months behind & 61.35 (40.10) & 17.09 (10.50) & 31.50 (23.57) & 57.34 (31.20) & 25.50 (23.57) & 25.45 (10.50)\\
        Wget & avoidable hi+ vulnerabilities & 7.00 (7.00) & 4.60 (7.00) & 5.18 (6.00) & 7.00 (7.00) & 4.44 (4.00) & 4.55 (6.00) \\
         & versions behind & 18.40 (17.00) & 12.40 (18.00) & 12.53 (12.00) & 25.00 (25.50) & 11.47 (10.00) & 13.00 (16.00)\\
         & months behind & 73.93 (63.40) & 57.75 (79.70) & 44.03 (41.23) & 119.69 (123.55) & 40.64 (34.33) & 55.47 (54.03)\\
        OkHttp & avoidable vulnerabilities & 0.82 (1.00) & 0.65 (1.00) & NaN & 0.38 (0.00) & 0.63 (1.00) & 0.75 (1.00)\\
         & versions behind & 25.09 (25.00) & 27.38 (28.00) & NaN & 22.50 (17.50) & 19.82 (19.00) & 18.32 (16.00) \\
         & months behind & 32.87 (38.60) & 34.24 (32.67) & NaN & 15.11 (9.97) & 20.04 (17.25) & 23.26 (19.47) \\
        Requests & avoidable vulnerabilities & 0.50 (0.50) & 0.75 (1.00) & 0.67 (1.00) & 1.95 (2.00) & 0.18 (0.00) & 0.22 (0.00)\\
         & versions behind & 16.00 (16.00) & 21.42 (27.50) & 15.33 (5.00) & 45.13 (46.00) & 4.58 (0.00) & 6.72 (1.00) \\
         & months behind & 24.17 (24.17) & 30.63 (37.47) & 21.72 (16.07) & 48.15 (48.27) & 7.61 (0.00) & 9.76 (1.42) \\
    \bottomrule
    \end{tabular}
    \label{table:cat-stats}
\end{table*}

\section{Plots}
\subsection{Avoidable vulnerabilities vs versions behind for all kinds of devices}\label{remaining-avoidable-plots}
Figures in the Appendix \ref{remaining-avoidable-plots} shows the plots for avoidable vulnerabilities vs versions behind for Wget, OkHttp, and Python Requests in Figure \ref{fig:okhttp-vuln-version}. The number of avoidable vulnerabilities and CDF of Wget, OkHttp, and Python request shows similar trend to Curl, see Figure \ref{fig:curl-vuln-version}. The shape of OkHttp and Requests appear to be different because they have fewer vulnerabilities, and that results in zoomed in Y axis; possibly because OkHttp and Requests are written in memory safe languages Java, and Python respectively. Otherwise, OkHttp and Requests also show a long tail distribution. To list the software components in order form most vulnerable to secure, we could look at number of avoidable vulnerabilities in those components, and from that Curl seems most vulnerable, then Wget, Python Requests, and finally OkHttp.

By looking at the plots of Python Requests and OkHttp, one could make an argument that if software component has less vulnerabilities, using newer versions doesn't result in improved the security, but those few vulnerabilities in older version could be exploited by attackers and do as much damage as having large number of vulnerabilities. Unless a software has no vulnerabilities, it's always beneficial to use the latest version.

\subsubsection{Plot of Wget, OkHttp, and Python Requests is in Figure \ref{fig:okhttp-vuln-version}}

\subsection{Distribution of number of avoidable vulnerabilities, number of versions behind, and number of month behind}\label{remaining-distro-plots}
Figures in the \ref{remaining-distro-plots} shows the distribution of distribution of number of avoidable vulnerabilities, number of versions behind, and number of month behind for different categories for Wget, and Python Requests in Figure \ref{fig:wget-cat}. From both the plots we can see that distribution for different categories are different. Looking at the mean and median, update practices of computing group is better than both IoT non-platforms and IoT platforms.
\subsubsection{Plot of Wget and Python Requests in Figure \ref{fig:wget-cat}}

\subsection{Plots for versions of software components used by vendors over our capture duration}\label{remaining-ttr}
Figures in the appendix \ref{remaining-ttr} depicts the version of Python Requests used by vendor eero Wap. Similar to Figure \ref{fig:curl-ttr} for cURL and OkHttp, we see that vendor has rolled out an updates to versions 2.21.0 and 2.23.0 and these updates are 1 and 3 versions behind from latest available at that time respectively.

We have not included the plots for Wget because we did not see any vendor updating Wget in our data.
\subsubsection{Plot of Python Requests in Figure \ref{fig:req-ttr}}

\section{Statistical values of metrics for all software components} \label{cat-stats-appendix}
Table \ref{table:cat-stats} lists the mean and median values of metrics used by us.

\end{document}